\documentclass[aps,pre,showpacs,floats,onecolumn,preprint,floats,superscriptaddress,floatfix,longbibliography]{revtex4}

\usepackage{bm}
\usepackage{hyperref}
\usepackage[usenames,dvipsnames]{xcolor}
\usepackage[normalem]{ulem}
\usepackage[utf8]{inputenc}
\usepackage{amsmath}
\usepackage{amsthm}
\usepackage{amsmath}
\usepackage{amsfonts}
\usepackage{tabularx}
\usepackage{tabulary}
\usepackage{blindtext}
\usepackage{booktabs}
\usepackage[utf8]{inputenc}
\usepackage[T1]{fontenc}
\usepackage{graphicx}
\usepackage{float}
\usepackage{wrapfig}
\usepackage{bm}
\usepackage{braket}
\usepackage{grffile}
\usepackage[normalem]{ulem}
\usepackage[usenames,dvipsnames]{xcolor}

\begin{document}

\title{DeepONet prediction of linear instability waves in high-speed boundary layers}

\author{Patricio Clark Di Leoni}
\affiliation{\small Department of Mechanical Engineering, Johns Hopkins University, Baltimore, MD 21218, USA}
\author{Lu Lu}
\affiliation{\small Department of Chemical and Biomolecular Engineering, University of Pennsylvania, Philadelphia, PA 19104, USA}
\author{Charles Meneveau}
\affiliation{\small Department of Mechanical Engineering, Johns Hopkins University, Baltimore, MD 21218, USA}
\author{George Karniadakis}
\affiliation{\small Division of Applied Mathematics and School of Engineering, Brown University, Providence, RI 02912, USA}
\author{Tamer A.~Zaki}
\thanks{corresponding author, email: \url{t.zaki@jhu.edu}}
\affiliation{\small Department of Mechanical Engineering, Johns Hopkins University, Baltimore, MD 21218, USA}

\begin{abstract}
Deep operator networks (DeepONets) are trained to predict the linear amplification of instability waves in high-speed boundary layers and to perform data assimilation. In contrast to traditional networks that approximate functions, DeepONets are designed to approximate operators. Using this framework, we train a DeepONet to take as inputs an upstream disturbance and a downstream location of interest, and to provide as output the perturbation field downstream in the boundary layer. DeepONet thus approximates the linearized and parabolized Navier-Stokes operator for this flow. Once trained, the network can perform predictions of the downstream flow for a wide variety of inflow conditions, without the need to calculate the whole trajectory of the perturbations, and at a very small computational cost compared to discretization of the original equations. In addition, we show that DeepONets can solve the inverse problem, where downstream wall measurements are adopted as input and a trained network can predict the upstream disturbances that led to these observations.  This capability, along with the forward predictions, allows us to perform a full data assimilation cycle: starting from wall-pressure data, we predict the upstream disturbance using the inverse DeepONet and its evolution using the forward DeepONet. 
\end{abstract}

\maketitle

\section{Introduction}
\label{sec:tbl}

\begin{figure}[htbp]
    \includegraphics[width=0.9\textwidth]{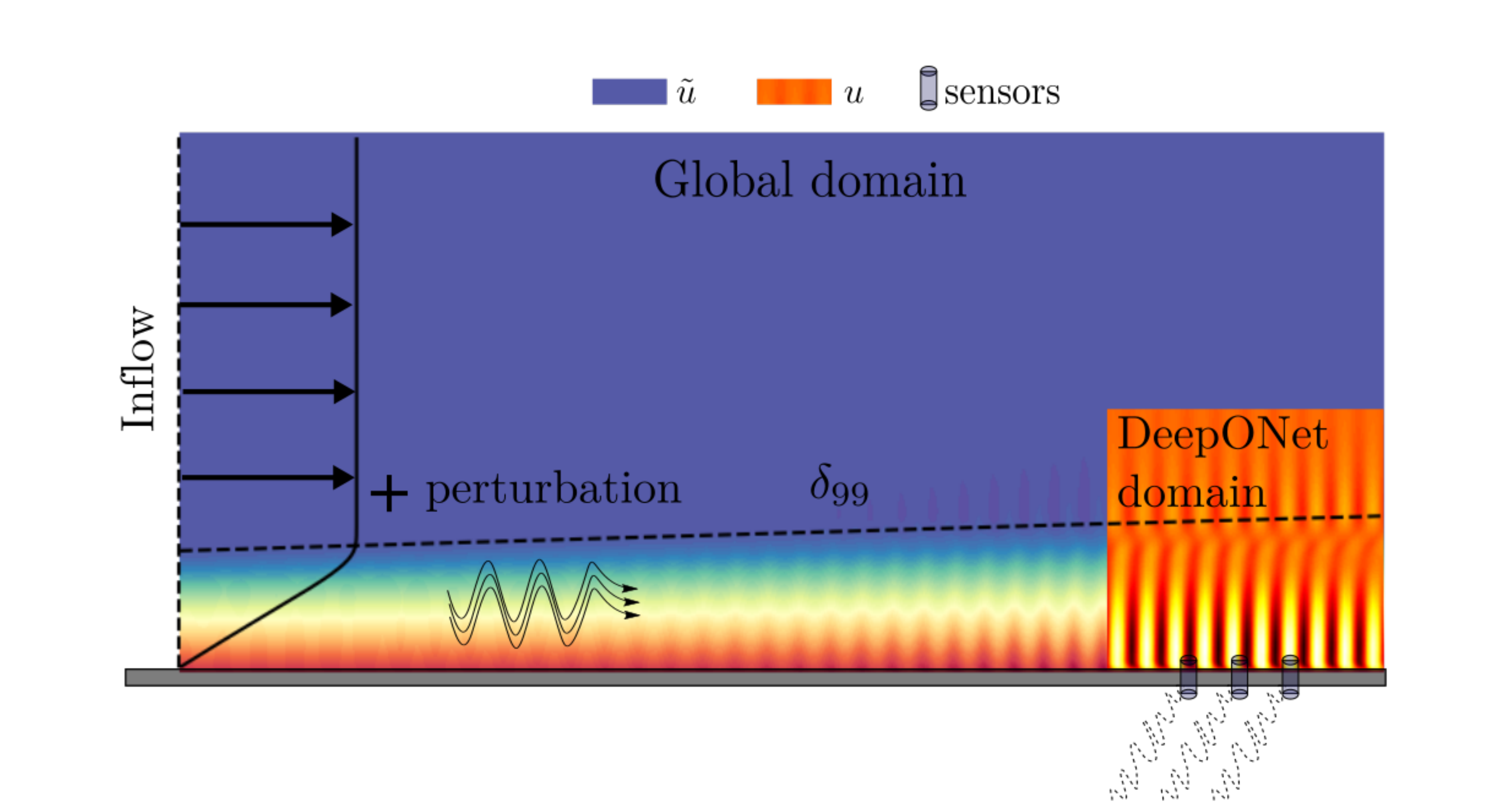}
    \caption{Visualization of an instability wave in a spatially developing boundary layer. At the inlet to the computational domain, the base flow is superposed with instability waves.  The dashed line marks the 99\% thickness of the boundary layer. The objective is to accurately predict the downstream evolution of the instability wave. }
    \label{fig:developing_bl}
\end{figure}

The early stages of transition to turbulence in high-speed flight often involve the exponential amplification of linear instability waves, which ultimately become nonlinear and cause breakdown to turbulence. 
The potential impact of premature transition on a flight vehicle can be undesirable due to the increased drag or even catastrophic due to the excessive local heating.
Therefore, ongoing research aims to accurately model each stage of the transition process using theory \cite{Mack1984,fedorov_transition_2011}, simulations \cite{zhong_direct_2012} and experiments \cite{schneider_developing_2015}.  
At every stage, the transition process can be significantly altered by uncertain elements, e.g., in the flow profile and boundary conditions \cite{park2019sensitivity},  free-stream noise \cite{schneider_effects_2001,joo_continuous_2012} or vibrations \cite{frendi_coupling_1993}.  While nonlinear optimization  strategies can be adopted to discover the most dangerous configurations and to mitigate them \cite{jahanbakhshi2019nonlinearly,jahanbakhshi2021control}, these approaches are computationally very costly.  Accurate and efficient approaches to predict the different stages of transition, starting with the early development of exponential instability waves, which is the focus of the present effort, are therefore pacing items for robust design and optimization of high-speed flight \cite{Leyva2017,Vision_2030_Final_Report_CRDL}. Figure~\ref{fig:developing_bl} shows a visualization of an instability wave in a high-speed, spatially developing boundary layer: the goal is to accurately predict  how the upstream instability wave will amplify or decay within a region of interest downstream. 

Several data-driven methods have been proposed to determine the amplification of instability waves, ranging from complex data fits \cite{drela_viscous-inviscid_1987, perraud_stability-based_2016} and numerical look-up tables \cite{krumbein_en_2008, pinna_reduced_nodate, saint-james_database_2020} to artificial neural networks \cite{crouch_transition_2002, fuller_neural_nodate, danvin_laminar_nodate, zafar_convolutional_2020}. All these approaches attempt to predict amplification factors over a range of frequencies of the instability waves, Reynolds numbers and flow conditions based on data generated from a stability theory and $e^N$ method. One issue that all these methods encounter is how to account for the shape of incoming perturbations. One commonly adopted approach is to reduce its functional form to a single number, namely the shape factor, and use this value as input to the prediction method. This approach has been shown to lack the necessary expressivity, and hence alternatives that rely on generating reduced-order representations using convolutional neural networks have been developed \cite{zafar_convolutional_2020}, although the need to accommodate functional input still remains. 

Of equal importance and complexity is the inverse of the above-described problem, namely determining incoming perturbations from downstream data.  This class of problems falls within the realm of data assimilation (DA). In the context of fluid dynamics, DA has found success in tackling the problem of state reconstruction using a variety of techniques, such as adjoint methods \cite{Wang_adjoint_2019,wang_spatial_2019,wang_state_2020}, ensemble approaches \cite{mons_kriging-enhanced_2019}, nudging \cite{foias_discrete_2016,clark_di_leoni_synchronization_2020} and neural networks \cite{raissi_hidden_2018,jin_nsfnets_2020,buzzicotti_reconstruction_2020}. In the context of high-speed boundary-layer stability,  \cite{buchta_2021} utilized an EnVar method to determine inflow perturbations from downstream wall-pressure measurements.  The sensors in that case where placed in the transitional and fully turbulent regions, and the solution of the linear problem demonstrated the value of these techniques.  One important consideration is that the EnVar procedure must be repeated for each new set of measurements.  Therefore, pre-trained neural networks, which can be evaluated very quickly relative to performing a new simulations, have the potential to accelerate the solution of these inverse problems.  

Traditional neural networks, by virtue of the universal approximator theorem \cite{cybenko_approximation_1989}, are built to approximate functions. While useful for many applications, they may be at a disadvantage when tackling the problems described above where the input is itself a function, e.g., the shape of the upstream instability wave. 
Two approaches can deal with this type of input. 
The first is the notion of an evolutional deep neural network (EDNN~\citep{du2021ednn}), which can solve the governing equations to evaluate the downstream evolution of disturbances. In this framework, the network is akin to a basis function and its initial state represents the upstream disturbance; the network parameters are subsequently updated using the governing equations to predict the downstream evolution of the instability waves \citep{du2021ednn}. This approach is therefore accurate, predictive and solves the governing equations for every new configuration.   

Here we focus on a different class of networks that rely on offline training to learn the operator that governs the dynamics of the instability waves and, once trained, can make fast prediction. Such networks harness the power of the universal operator approximator theorem \cite{chen1995universal}, which states that neural networks can approximate functional and operators with arbitrary accuracy, and have shown great promise. For example, \cite{ferrandis_learning_2019} utilized long short-term memory (LSTM) networks to learn a functional that predicts the motion of vessels in extreme sea states, and \cite{li_fourier_2020} developed a Fourier-based method and used it to predict the evolution of 2D incompressible flows. Of particular importance to our work is the deep operator network (DeepONet, \cite{lu_learning_2021}), a branched type of neural network based directly on the theorems of \cite{chen1995universal} that has proven able to successfully model a wide range of problems including ODEs, PDEs and fractional differential operators. Beyond the cost of their training, these networks are able to approximate the targeted operators with high accuracy and speed. DeepONets can also be easily integrated into data assimilation schemes involving multiple physics and multiple scales; for example,  \cite{cai_deepmmnet_2021} applied them to electroconvection and \cite{mao_deepmmnet_2020} demonstrated their ability to predict the flow and finite-rate chemistry behind a normal shock in high-speed flow.

In this context, we propose to use DeepONets for
the prediction of the evolution of instability waves in transitional
boundary layers in compressible flows. We show that DeepONets can learn
to reproduce solutions of the parabolized stability equations (PSE), a
linearized and parabolized set of equations derived from the full
Navier-Stokes equations that describe the evolution of perturbations in a
developing boundary layer. Moreover, we demonstrate how DeepONets
can be used to tackle the inverse problem of determining the upstream
disturbance environment from limited wall measurements. These results
open new and promising avenues for data assimilation and control of
boundary layer flows.

The paper is organized as follows. In Section~\ref{sec:deeponets} we present the DeepONet architecture and provide a simple example. In Section~\ref{sec:tbl_eqs} we outline the equations and regions in parameters space of the flow, while in Section~\ref{sec:data_generation} we explain the details of the data generation and of the training protocols. The results are presented in Section~\ref{sec:results} and the conclusions in Section~\ref{sec:conclusions}.

\section{DeepONet architecture}
\label{sec:deeponets}

\begin{figure}[htbp]
    \includegraphics[width=0.55\textwidth]{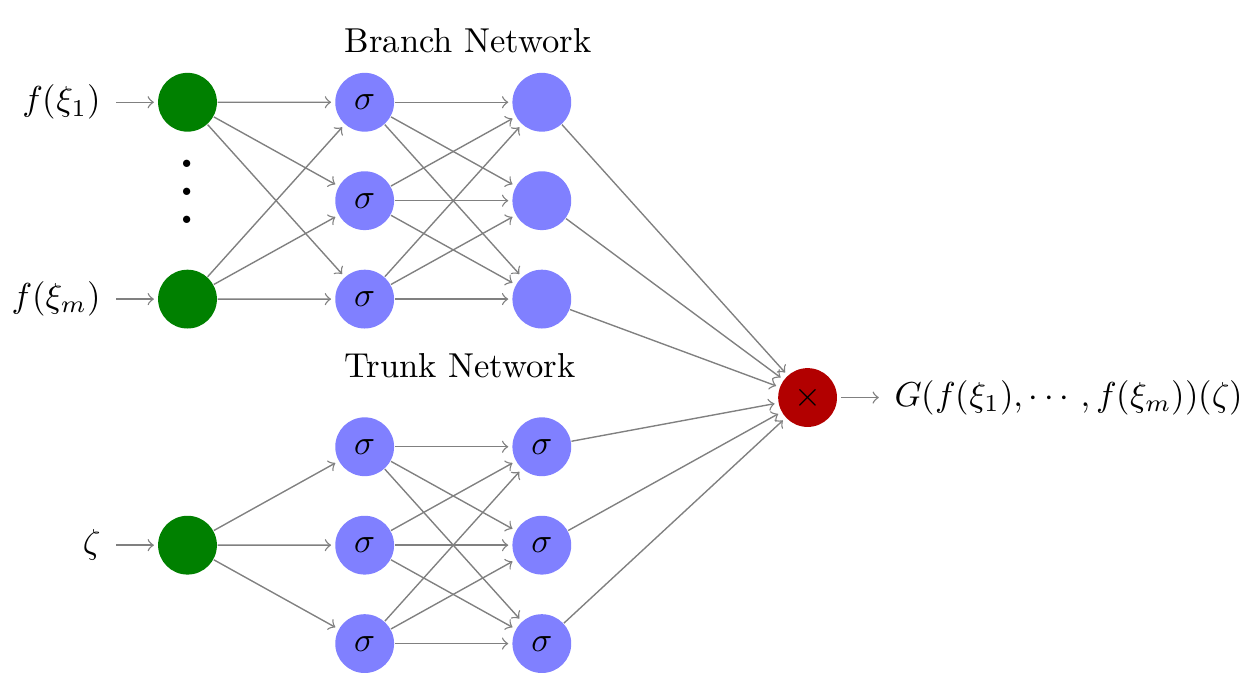}
    \caption{The DeepONet architecture. The green nodes on the left indicate the input notes. The blue nodes indicate the hidden units, with the ones marked with $\sigma$ denoting those that use activation function $\sigma$. The crossed red node on the right indicates the output, which is obtained by taking the dot product of the final layers for the branch and trunk networks.}
    \label{fig:tbl_deeponet}
\end{figure}

Here we present some background on DeepONet, which follows the original presentation in \cite{lu_learning_2021}. Let $G^\dagger$ be an operator which maps an input function $f$ to an output function $G^\dagger (f)$, and let $\zeta \in Y$ be a point in the domain of the output function ($Y$ can be a subset of $\mathbb{R}$ or $\mathbb{R}^n$, indistinctly). We define points $[\xi_1, \xi_2, \cdots, \xi_m]$ in the domain of $f$, such that $[f(\xi_1), f(\xi_2), \cdots, f(\xi_m)]$ is a discrete representation of $f$.
The objective of a DeepONet is to approximate the operator $G^\dagger(f)(\zeta)$ by a DeepONet $G(f(\xi_1), f(\xi_2), \cdots, f(\xi_m))(\zeta)$. The structure of $G$ is shown in Figure~\ref{fig:tbl_deeponet} and corresponds to the ``stacked'' version presented in \cite{lu_learning_2021}. The input is indicated with green nodes, hidden units are indicated with blue nodes, and the output is indicated with a red node. The network is separated into two subnetworks, a branch network of depth $d_b$ that handles the discretized function input, and a trunk network of depth $d_t$ that handles the input of the final function. Each subnetwork consists of a fully-connected feed-forward neural network where every hidden unit is passed through an activation function $\sigma$, except for the last layer of the branch network which does not go through any activation function. The depth or number of layers of both subnetworks can be different, and similarly the widths of their different layers, apart from their respective last layers which have to be of equal width. In practice, we use the same number of hidden units $p$ in every layer. The final output is obtained by performing the dot product of the last layer of each subnetwork plus a bias term, i.e.
\begin{equation}
    G(f(\xi_1), f(\xi_2), \cdots, f(x_m))(\zeta) = \sum_{k=1}^{p}b_{k}t_{k}+b_{0},
\end{equation}
where $b_k$ and $t_k$ are the values of the hidden units for the last layer of the branch and trunk networks and $b_0$ is an extra bias term. For simplicity, we omit the explicit discretization of the functional input in $G(f(\xi_1), f(\xi_2), \cdots, f(\xi_m))(\zeta)$ from now on and abbreviate it as $G(f)(\zeta)$. Finally, DeepONets can be trained by minimizing a loss function of the type
\begin{equation}
    L = \frac{1}{N} \sum_{i=1}^N w_i \vert G(f_i)(\zeta_i) - G^\dagger(f_i)(\zeta_i) \vert^2,
    \label{eq:loss_deeponet}
\end{equation}
where $(f_i, \zeta_i)$ are the $N$ different pairs of functions and trunk inputs used for training and $w_i$ is the associated weight, which in the simplest case is taken to be equal to unity for every sample. The training can be performed with either a single batch high order method like L-BFGS or a mini-batch based stochastic gradient descent method like Adam. A comparison between DeepONets and convolutional neural networks (CNNs) is presented in Appendix~\ref{app:cnns}.

\subsection{A simple example}

As an illustrative example of DeepONets, we apply it to the anti-derivative operator similar to \cite{lu_learning_2021}. Given  
\begin{equation}
    \frac{d g(\zeta)}{d\zeta} = f(\zeta),
\label{eq:derivative}
\end{equation}
with $g(0)=0$ and $\zeta \in Y = [0,1]$, the goal is to approximate 
\begin{equation}
    G^\dagger (f)(\zeta) = \int^\zeta_0 f(\xi) d\xi.
\label{eq:anti_op}
\end{equation}
In order to train a DeepONet to evaluate $G$, one first needs to select $m$ points $[\xi_1, \xi_2, \cdots, \xi_m]$ on the domain $\xi \in [0,1]$, generate a training dataset using $N$ pairs of functions $f_i$ and points $\zeta_i$ and solve Eq.~\eqref{eq:derivative}. While in principle the functions $f$ could be any kind of function, restricting them to a certain function class, like polynomials or Gaussian random fields, will ease the training process. Both the functions $f_i$ and the points $\zeta_i$ can be repeated throughout the dataset, i.e., one can evaluate the same $G(f)$ at different locations or also using the same location for different $G(f)$. Once the dataset is generated, the DeepONet can be trained by minimizing the loss function \eqref{eq:loss_deeponet}. The trained DeepONet will then approximate $G(f)(\zeta)$ for any $f$ inside the function class used and for any $\zeta$ in the training domain $Y$.

\begin{figure}[htbp]
    \includegraphics[width=0.32\textwidth]{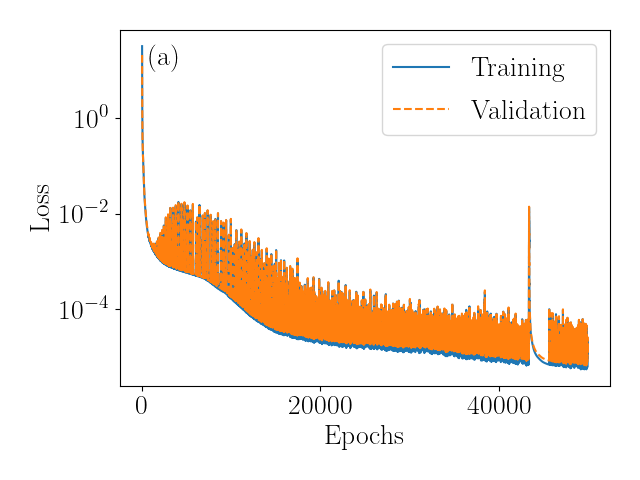}
    \includegraphics[width=0.32\textwidth]{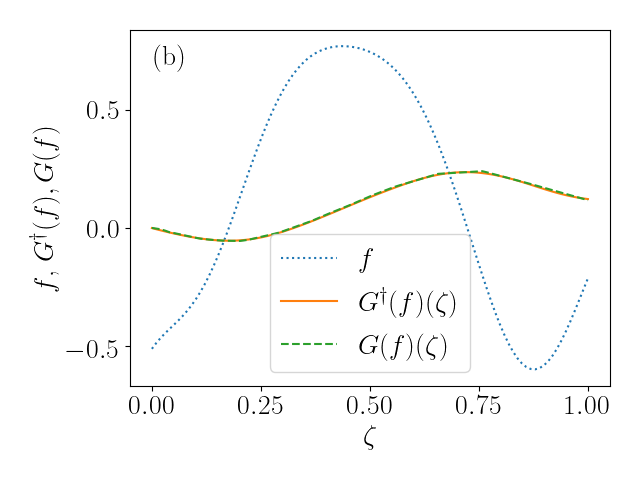}
    \includegraphics[width=0.32\textwidth]{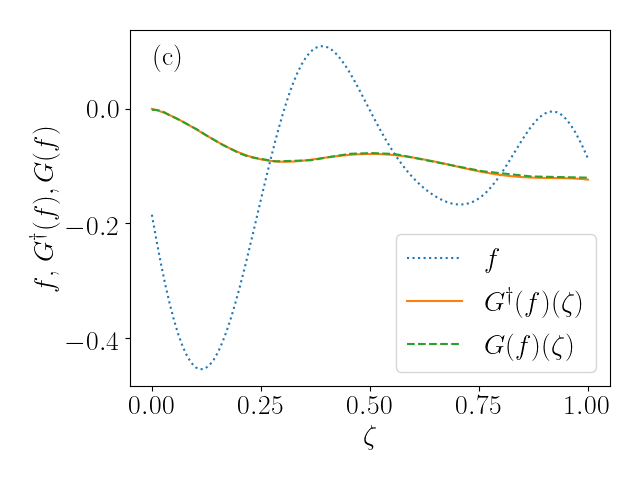}
    \caption{(a): Evolution of the loss function evaluated on the training and validation datasets for the DeepONet trained to learn the antiderivative operator in Eq. \eqref{eq:anti_op}.
    (b) and (c): Two different examples extracted from the validation dataset for the DeepONet trained to learn the anti-derivative operator evaluated over the whole domain $Y$. The input function $f$ is plotted along the data $G^\dagger(f)(\zeta)$ and the DeepONet prediction $G(f)(\zeta)$.}
    \label{fig:loss_anti}
\end{figure}

We implemented this example using Gaussian random fields, which are generated by a Gaussian process with a radial basis function kernel (which imposes Gaussian correlations) with a fixed correlation length $\ell=0.2$, as the function class. The training and validation datasets were each comprised of $10,000$ different functions, each evaluated at a single $\zeta$ location, but different from each other. The architecture of the DeepONet was $d_b=d_t=2$, $p=40$ and $m=100$, and the network was optimized using an Adam optimizer with learning rate $\eta=10^{-3}$. In Figure~\ref{fig:loss_anti}(a) we show the evolution of the loss evaluated on the training and validation datasets as a function of the number of epochs; the network is able to minimize the loss without any overfitting. In Figures~\ref{fig:loss_anti}(b) and (c) we show two examples from the validation set. The DeepONet prediction $G(f)(\zeta)$ matches $G^\dagger(f)(\zeta)$ visibly well, with mean local deviations below $3\%$ for both cases (b) and (c).

\section{Linear instability waves in compressible boundary layers}
\label{sec:tbl_eqs}

We now introduce the governing equations for high-Mach number flows, with our interest being the linear evolution of instability waves in zero-pressure-gradient boundary layers.
We take $(x,y)$ to be the streamwise and wall-normal coordinates. The state vector $\tilde{\mathbf{q}} = (\tilde{{\rho}}, \tilde{{u}}, \tilde{{v}}, \tilde{{T}})$ is comprised of the fluid density, the velocity components in the two coordinate directions and the temperature.
The free-stream has characteristic velocity $U_0$, temperature $T_0$, specific heat ratio $\gamma_0$, viscosity $\mu_0$ and density $\rho_0$.  The starting location of the flow domain under consideration is located at $x_0$, and the Blasius length $L_0 = \sqrt{\mu_0 x_0/\rho_0 U_0}$ is adopted as the characteristic lengthscale.  The inflow Reynolds number is therefore $Re_0 = \rho_0 U_0 L_0 /\mu_0$ and the Mach number is $M_0 = U_0 / \sqrt{\gamma_0 \mathcal{R} T_0}$ where $\mathcal{R}$ is the gas constant. 

The flow satisfies the Navier-Stokes equations for an ideal compressible gas,
\begin{gather}
\frac{\partial \tilde{\rho}}{\partial t}+\bm{\nabla} \cdot(\tilde{\rho} \tilde{\bm{u}})=0,
\\
\frac{\partial \tilde{\rho} \tilde{\bm{u}}}{\partial t}+\bm{\nabla} \cdot(\tilde{\rho} \tilde{\bm{u}} \tilde{\bm{u}}+\tilde{p} \bm{I}-\bm{\tau})=0,
\\
\frac{\partial E}{\partial t}+\bm{\nabla} \cdot(\bm{u}[E+\tilde{p}]+\bm{\theta}-\tilde{\bm{u}} \cdot \bm{\tau})=0,
\end{gather}
where $\tilde{\bm{u}}$ is the velocity vector, $\bm{I}$ is the unit tensor, $E=\tilde{\rho} e + 0.5 \tilde{\rho}\tilde{\bm{u}}\cdot \tilde{\bm{u}}$ is the total energy, $e$ is the specific internal energy, $\bm{\tau}$ is the viscous stress tensor, and \bm{$\theta$} is the heat-flux vector. Thermodynamic relations for $\tilde{p}$ and $\tilde{T}$, as well as the expression for $\bm{\tau}$ and $\bm{\theta}$ close the system and can be found in the literature, e.g., \cite{jahanbakhshi2019nonlinearly}.
    
\subsection{Parabolized Stability Equations}

In the early stages of their development, small-amplitude instability waves in a boundary layer can be accurately described by the linear parabolized stability equations (PSE).  The equations are derived from the Navier-Stokes equations by decomposing the flow state $\tilde{\mathbf{q}}$ into the sum of a base flow and a perturbation (see Fig.\,\ref{fig:developing_bl}).  The base flow in this case is the undistorted, spatially developing boundary-layer solution $\bm{Q} = (\rho_B, U_B, V_B, 0, T_B)^T$.  The equations governing the perturbation field $\bm{q} = ({\rho}, {u}, {v},  {T})^T$ are then linearized,  
\begin{equation}
    \mathcal{V}_{t}\frac{\partial \textbf{q}}{\partial t} 
    +\textbf{L}(\textbf{Q})\textbf{q}=0,
    \label{eq:lin_ptb_eq}
\end{equation}
where $\mathcal{V}_{t}$ is the linear operator matrix, and $\textbf{L}$ is the linear differential operator matrix \cite{chang1993linear} 
\begin{align*}
    \textbf{L} = &\mathcal{V}_{0}+\mathcal{V}_{x}\frac{\partial}{\partial x} 
    +\mathcal{V}_{y}\frac{\partial}{\partial y}+
    \\
    &\mathcal{V}_{xx}\frac{\partial^{2}}{\partial x^{2}}
    +\mathcal{V}_{xy}\frac{\partial^{2}}{\partial x \partial y}
    +
    \mathcal{V}_{yy}\frac{\partial^{2}}{\partial y^{2}}.
\end{align*}
The exact form of the operator matrices $\mathcal{V}$ is provided in
Appendix~\ref{appendix:matrix}. We introduce the following ansatz for the perturbations,
\begin{equation}
    \bm{q} = \check{\bm{q}}(x,y) \exp \left( \int_{x_0}^{x} \alpha(s) ds - i \omega t + i \phi\right) + c.c., 
    \label{eq:pse_decomp}
\end{equation}
where $\check{\bm{q}} = (\check{\rho}, \check{u}, \check{v}, 
\hat{T})^T$, $\alpha$ is the local complex-value streamwise wavenumber,
$\omega$ is the perturbation
frequency, and $\phi$ is the phase. Substituting this expression into
Eq.~\eqref{eq:lin_ptb_eq} yields the PSE, 
\begin{equation}
    \check{\mathcal{A}} (\bm{Q}) \frac{\partial \check{\bm{q}}}{\partial x}
    + \mathcal{L}(\bm{Q}, \alpha, \omega) \check{\bm{q}} = 0,
    \label{eq:pse}
\end{equation}
where $\check{\mathcal{A}}$ and $\mathcal{L}$ are linear differential operators whose expressions are provided in Appendix~\ref{appendix:matrix}.  Further details and explanations about the derivation of the PSE can be found in Refs.~\cite{chang1993linear,park2019sensitivity}.  This formulation is linear and allows for the problem to be marched downstream, instead of solving the full domain all at once. Nonetheless, the solution procedure requires careful consideration to ensure numerical stability and accuracy.  Once solved, the generated data can be used to train neural networks that are better suited for making fast predictions. 

\subsection{Unstable modes}
\label{sec:unstable_modes}

\begin{figure}
    \includegraphics[height=0.5\textwidth]{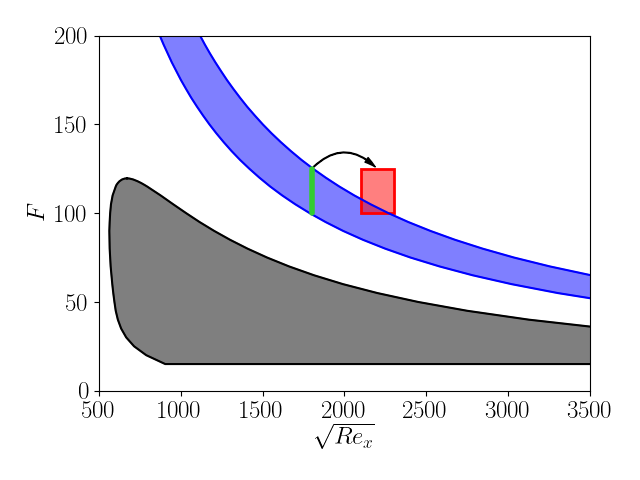}
    \caption{Neutral stability curves of the compressible boundary layer at $Ma=4.5$. The gray and blue regions mark the first and second unstable Mack modes, respectively. The green line denotes the region of parameter space used as input for the DeepONet, while the red square denotes the output region.}
    \label{fig:neutral_curves}
\end{figure}

The disturbances of interest are instability waves, which depend on the flow parameters.  We will consider air with Prandtl number $Pr = 0.72$ and ratio of specific heats $\gamma = 1.4$.  The free-stream Mach number is $Ma=4.5$ and the free-stream temperature is $T_0=65.15K$.  The neutral curves for spatial instabilities in a zero-pressure-gradient boundary layer are shown in Figure~\ref{fig:neutral_curves}, reproduced from \cite{jahanbakhshi2019nonlinearly}.  The instability frequency is $F= \omega 10^6 / \sqrt{Re_{x_0}}$ and the local Reynolds number is $\sqrt{Re_{x_0}}$. The two shaded regions mark two different classes on unstable modes: The lower region corresponds to three-dimensional vortical instabilities that have their origin in Tollmien-Schlichting waves when traced back to lower Mach numbers;  the upper region corresponds to the Mack second modes.  At our Mach number and above, the Mack modes start to become dominant and are recognized to be a key contributor to transition at high Mach numbers.  
The transition Reynolds number observed in high-altitude flight test at $Ma>4$ is usually greater than $\sqrt{Re_{x_0}}=2000$ \cite{harvey_influence_1978, schneider_flight_1999}.  For this reason we set the inflow location of our configuration slightly upstream, at $\sqrt{Re_{x_0}}=1800$.  At this chosen inflow Reynolds number, the unstable two-dimensional Mack second modes span the frequency range $100 \lesssim F \lesssim 125$ as shown in Figure~\ref{fig:neutral_curves}. This frequency range will be the focus of our DeepONet training.  


\section{problem set-up and dataset generation}
\label{sec:data_generation}

\begin{figure}
    \includegraphics[width=0.5\textwidth]{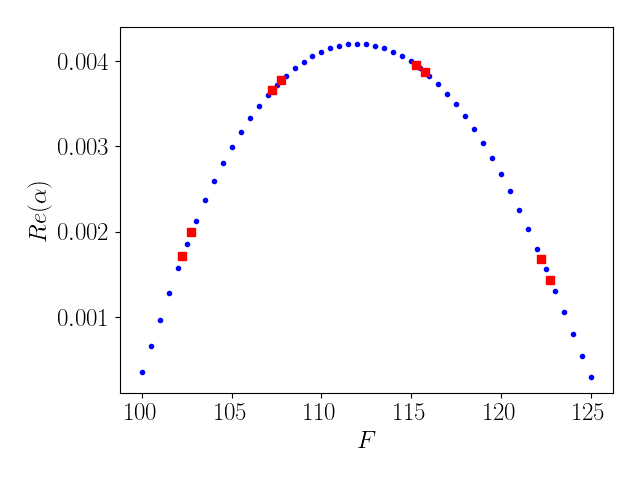}
    \caption{Real part of the streamwise wavenumber $\alpha$ as a
    function of perturbation frequency $F$. The modes marked with blue
    dots were used to generate the training dataset, while the modes
    marked with red squares were reserved for testing purposes.}
    \label{fig:alpha_vs_F}
\end{figure}

We generated data to train and test our DeepONets by simulating the evolution of instability waves using the parabolized stability equation \eqref{eq:pse} and the code described in \cite{park2019sensitivity}. The domain of integration spans $1800 \le \sqrt{Re_x} \le 2322$  and $y/L_0 \in [0, 220]$. The equations were solved for 59 different perturbation frequencies in the range $F\in[100,125]$. Figure~\ref{fig:alpha_vs_F} shows the real part of the streamwise wavenumber $\alpha$ as a function of the perturbation frequency $F$ that we consider. Modes marked with blue circles were used to generate the training datasets, while the eight modes marked with red squares were used as independent validation data. The resolution in frequency of the training dataset is $\Delta F= 0.5$. Within the range of $F$ of interest, only two-dimensional Mack modes were considered since they are recognized as an important precursor of transition in high-speed flows. Note that in addition to selecting the frequency of an instability wave, we can also arbitrarily adjust its phase\textemdash a property that we will exploit to augment the training data.

We separate the flow domain into three regions. The first, 
\begin{equation}
    Y_{u} = \{\sqrt{Re_x} = 1800,~y/L_0 \in [0,30],~t \in [0, 4T]\},
\end{equation}
is the {\it upstream} position where the instability waves enter the domain of interest and, at the position, are a function of the wall-normal coordinate and time. The second, 
\begin{equation}
    Y_{d} = \{\sqrt{Re_x} \in [2200, 2322],~y/L_0 \in [0,30],~t=0 \},
\end{equation}
is a {\it downstream} region where the instability waves depend on the streamwise and wall-normal position, and have potentially amplified or decayed relative to their inflow amplitude. The last region,
\begin{equation}
    Y_{w} = \{ \sqrt{Re_x} \in [2300, 2322],~y/L_0 = 0,~t \in [0, 4T] \},
\end{equation}
marks a narrow streamwise extent along the {\it wall}.  

We trained several DeepONets under a variety of setups. The first version is termed the {\it forward} case F, whose idea is illustrated  in Figure~\ref{fig:developing_bl}. In this case the goal is to map an inflow perturbation to its evolution downstream and we only work with the streamwise velocity component $u$ of the full solution $\bm{q}$. The input to the branch net is $u$ evaluated in the subdomain $Y_{u}$ discretized using 47 points in the wall-normal coordinate and 20 in time, thus totalling 940 sensors. The trunk net, on the other hand, is evaluated at a point downstream $(x,y) \in Y_{d}$. The output of the DeepONet is $u(x,y)$. Under the notation presented in Sec.~\ref{sec:deeponets}, $f=u$, $\xi = (y,t)\in Y_{u}$, $\zeta = (x,y)\in Y_d$ and $G(f) = u$. Note that the input and output domains are not adjacent.  The dataset for this case was generated by picking $N$ solutions with different frequencies $F$ (from the training set marked in Fig.~\ref{fig:alpha_vs_F}) and phases $\phi$. The goal of the DeepONet is shown schematically in Figure~\ref{fig:neutral_curves}, where the green line denotes the input and the the red area denotes the output region. From each solution only one point, chosen at random, was used for the trunk evaluation. The probability distribution used to sample the evaluation points favored the near-wall regions where most of the activity is concentrated. It is also possible to use a smaller pool of solutions and sample more points $(x,y)$ for the trunk evaluation; in our experience choosing either strategy yields similar prediction accuracy as long as the number of solutions with different frequencies and phases used is sufficiently large. 


The next cases focus on retrieving other field variables. Using the same setup as case F, we define cases F$_p$ and F$_T$, where the target outputs are now the pressure and temperature fields, respectively. Note that the input to both cases is still the streamwise velocity perturbation $u$ evaluated at $Y_{u}$. When solving the PSE, all five fields have to be evaluated concurrently, as they are coupled. With these two cases our objective is to show how a DeepONet, which learns an operator using data, may be trained separately for different fields.  DeepONet thus learns the PSE and the observation operator that extracts the specific field of interest.

\begin{figure*}[htbp]
    \includegraphics[width=0.75\textwidth]{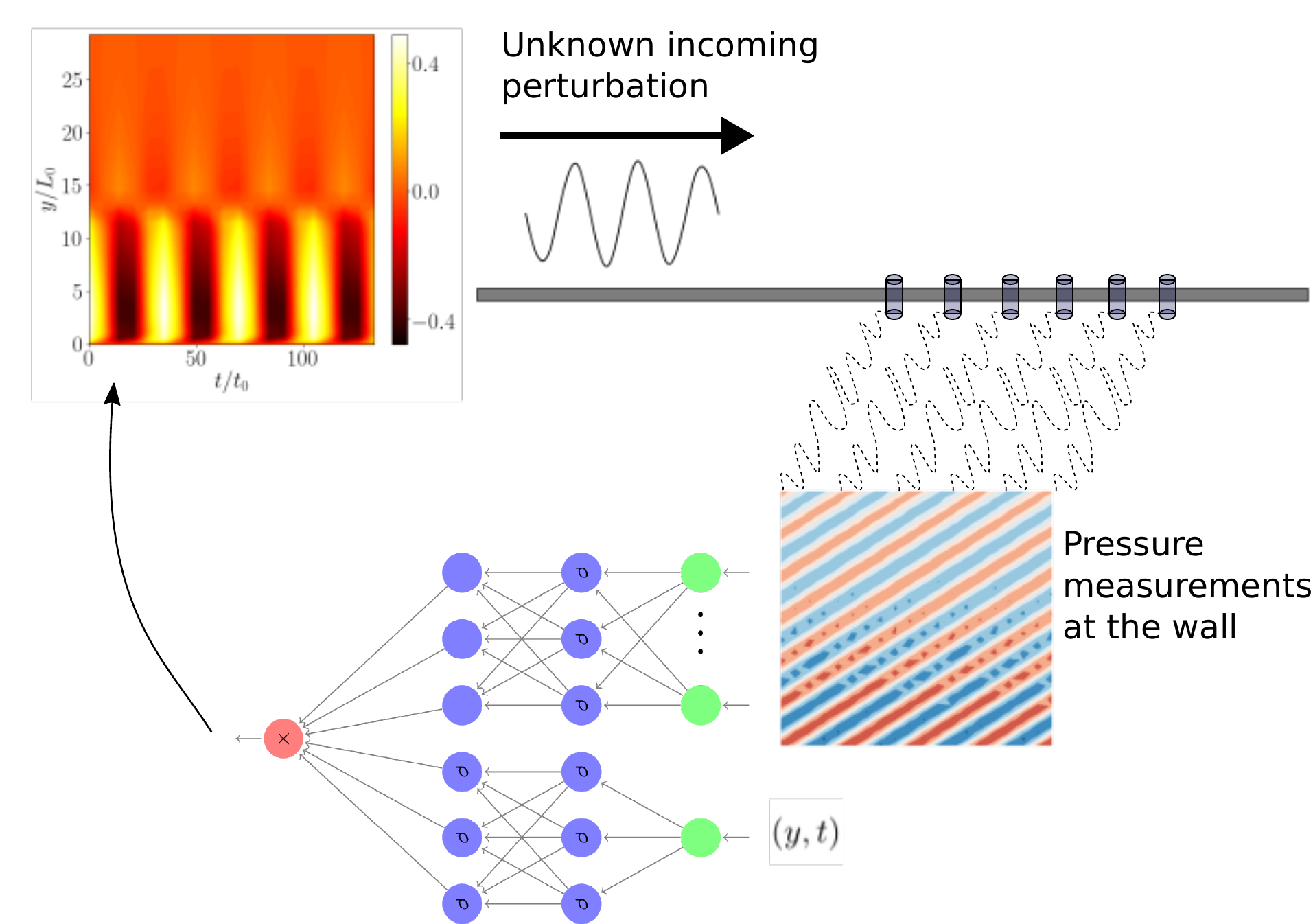}
    \caption{Diagram outlining the setup of the inverse problem. The
    DeepONet takes as input to the branch net measurements of the
    pressure at the wall downstream, and it outputs the inflow
    perturbation that generated such pressure fluctuations.}
    \label{fig:inverse_diagram}
\end{figure*}

As a counterpart to the forward case F, we define an {\it inverse} case I. The goal here is to reconstruct an inflow perturbation from downstream wall-pressure measurements. A diagram of this problem is shown in Fig.~\ref{fig:inverse_diagram}. We use the pressure field evaluated at $Y_{w}$ as input to the branch network, evaluate the trunk network at points $(y,t) \in Y_{u}$, and output $u$. The subdomain $Y_{w}$ is discretized using 47 points for $x$ and 40 points to $t$, totalling $1880$ sensors. 

Finally, as the PSE is a linear equation we can generate solutions with more than one perturbation frequency using the superposition principle. However, as neural networks are inherently non-linear operators, DeepONets cannot predict the evolution of superposed instability waves unless explicitly trained to do so. We therefore consider two further cases, F$_2$ and I$_2$, which expand upon cases F and I by considering pairs of instability waves, 
\begin{equation}
    \bm{q} = a_1 \bm{q}_1 + a_2 \bm{q}_2,
\end{equation}
where $\bm{q}_1$ and $\bm{q}_2$ are two different solutions and $a_1, a_2 \in [0.9, 1.1]$ their respective amplitudes, to generate the datasets. The training data in this case includes different frequencies and phases of the modes within a pair, as well as their amplitudes.

\subsection{Hyperparameters and training protocols}
\label{sec:hyper}

All cases were trained by minimizing the loss function Eq. \eqref{eq:loss_deeponet} using the Adam algorithm. Mini-batches of $1000$ elements and an initial learning rate $\eta=10^{-4}$ were used in every case. The learning rate was reduced to $10^{-5}$ if the value of the loss function reached a plateau or started to increase. An early stopping protocol was adopted in order to retain the optimal state. No further regularization procedures were adopted.
Due to the fact that some instability waves decay within the domain of interest, the fields of interest have amplitudes spanning at least two orders of magnitude. These imbalances can lead to difficulties during training, in particular low-amplitude waves become harder to learn, as will be shown below.  In order to mitigate this issue, we used $w_i = A^{-1}_i$ with
\begin{equation}
    A_i = \max_{\zeta\in Y} G(f_i)(\zeta),
    \label{eq:weights}
\end{equation}
as weights in the loss function in Eq.~\eqref{eq:loss_deeponet}. Only in two particular cases, case F$_{A0}$ and case F$_{A2}$, where $w_i =1$ and $w_i = A^{-2}_i$ respectively, were different weights used. Justifications for the various weightings tested will be provided in the next section.


\begin{table}[h]
\begin{center}
\begin{tabular}{ c | c | c | c | c}
 Cases & $d_b$ & $d_t$ & $p$ & $N$\\ 
 \hline
 F & 5 & 6 & 100 & $3.3 \times 10^6$
 \\
 F$_2$ & 6 & 6 & 20 & $2.6 \times 10^7$
 \\\
 I & 7 & 5 & 9 & $4.8 \times 10^6$
 \\\
 I$_2$ & 8 & 8 & 10 & $9.6 \times 10^6$
 \\
\end{tabular}
\end{center}
\caption{Depth of the branch network, $d_b$, depth of the trunk network, $d_t$, width of the networks, $p$ and number of elements in the training dataset used for each case, $N$. All forward cases use the same architecture as case F, except for F$_2$.}
\label{table:cases_hp}
\end{table}

The depth and width of the networks and the number of elements in the training dataset used is reported in Table~\ref{table:cases_hp}. In all cases, Exponential Linear Units (ELUs) were used as activation functions and the Glorot algorithm was used for initialization of the network. All networks also have an input and output min-max normalization layer that ensures the values entering and predicted by the network are between $-1$ and $1$.

The highly oscillating nature of the data can hinder convergence of the networks during training, with predictions being pinned at the mean value. To mitigate this potential difficulty, after the normalization layer we perform a harmonic feature expansion on the input of the trunk network, 
\begin{equation}
    \zeta \mapsto (\zeta, \cos(2^0 \pi \zeta), \sin(2^0 \pi \zeta), \cos(2^1 \pi \zeta), \sin(2^1 \pi \zeta), \cdots \cos(2^n \pi \zeta), \sin(2^n \pi \zeta)).
    \label{eq:feat1}
\end{equation}
All forward cases previously defined use feature expansion up to $n=1$ (i.e., including  wavenumbers $2^0 \pi$ and $2^1 \pi$). The inverse cases do not use features. To understand the impact of the feature expansion we define a last set of cases, all based on case F but with different number of features expansions: F$_{nf}$ with no harmonic features; F$_{n\{0,2,3,4\}}$ with features up to $n=\{0,2,3,4\}$.

As a summary, we present a list and short description of every case performed in Table~\ref{table:summary}.

\begin{table}[h]
\begin{center}
\begin{tabular}{ c | c }
 Cases & Description
 \\
 \hline
 F & Forward case
 \\
 F$_p$, F$_T$ & Forward cases mapping to pressure and temperature
 \\
 F$_{A0}$, F$_{A2}$ & Forward cases with alternative loss function weights
 \\
 F$_{n\{f,0,2,3,4\}}$ & Forward cases with different number of input features
 \\
 F$_2$ & Forward case with two-mode combinations
 \\\
 I & Inverse case
 \\\
 I$_2$ & Inverse case with two-mode combinations
 \\
 A & Data assimilation case
\end{tabular}
\end{center}
\caption{Summary of all the different cases presented. Details are provided in Section~\ref{sec:data_generation}.}
\label{table:summary}
\end{table}


\section{Results}
\label{sec:results}

\subsection{Forward problems}

\begin{figure}[htbp]
    \includegraphics[width=0.40\textwidth]{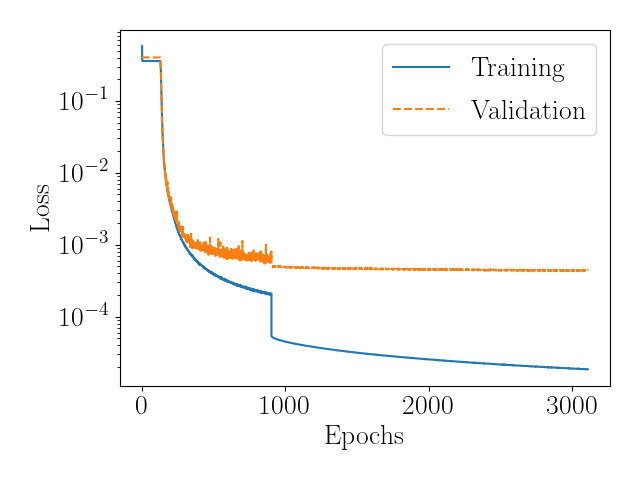}
    \caption{Evolution of the loss function evaluated on the training and validation datasets for case F.}
    \label{fig:loss_F}
\end{figure}

\begin{figure*}[htbp]
    \includegraphics[width=0.45\textwidth]{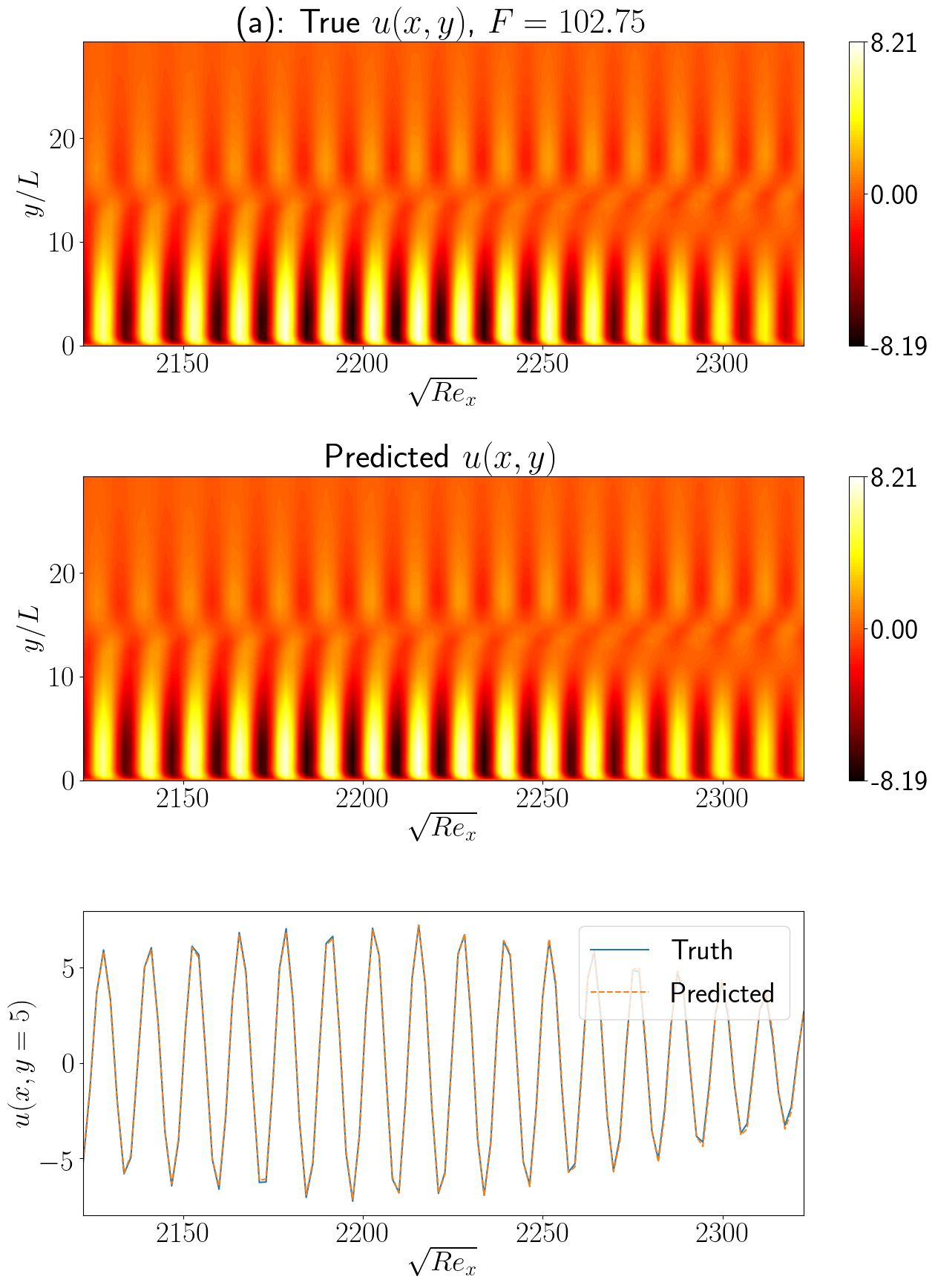}
    \includegraphics[width=0.45\textwidth]{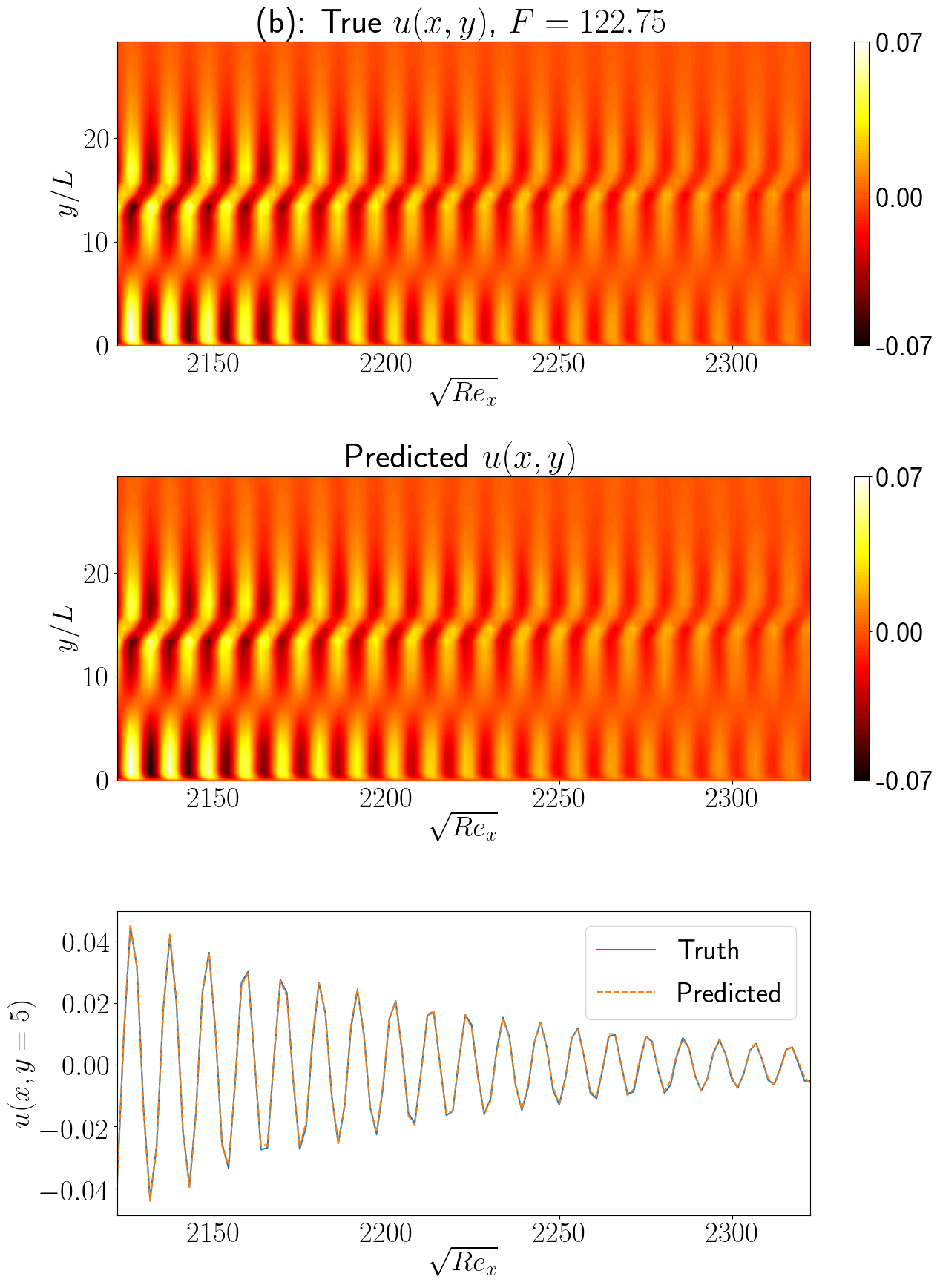}
    \caption{Examples from case F.
    For two particular input mode frequencies,
    (a) $F=102.75$ and (b) $F=122.75$, we show the true solutions as generated by the PSE on the top row, the prediction obtained from the DeepONet on the middle row, and a comparison of the profiles for a fixed $y/L_0=5$ on the bottom row. Both frequencies shown belong to the validation dataset.}
    \label{fig:examples_F}
\end{figure*}

We start by presenting the results of case F, where an inflow instability wave is mapped to the associated downstream velocity field. Figure~\ref{fig:loss_F} shows the evolution of the value of the loss function evaluated on the training and validation datasets as a function of the training epoch. After a brief plateau in the loss where the network outputs zero for every input (all solutions have zero mean), both curves decrease by several orders of magnitude which indicates that the DeepONet is able to learn the correct mapping of the data. As a first qualitative assessment of how well the trained DeepONet performs, we show in Figure~\ref{fig:examples_F} the prediction of two different modes evaluated over the whole output domain $Y_{d}$. The two modes were selected from the validation dataset, i.e.\,their frequencies were never seen during training. The figure also shows the true field $u(x,y)$ and a comparison of the profiles for a fixed $y/L_0=5$. The DeepONet correctly predicts the wall-normal profile, streamwise wavelength, phase, and amplitude of each solution. While the loss function takes much smaller values when evaluated on the training dataset compared to the validation data (Fig.\,\ref{fig:loss_F}), the apparent overfitting does not compromise the accuracy of prediction for modes within the validation set.

\begin{figure*}[htbp]
    \includegraphics[width=0.45\textwidth]{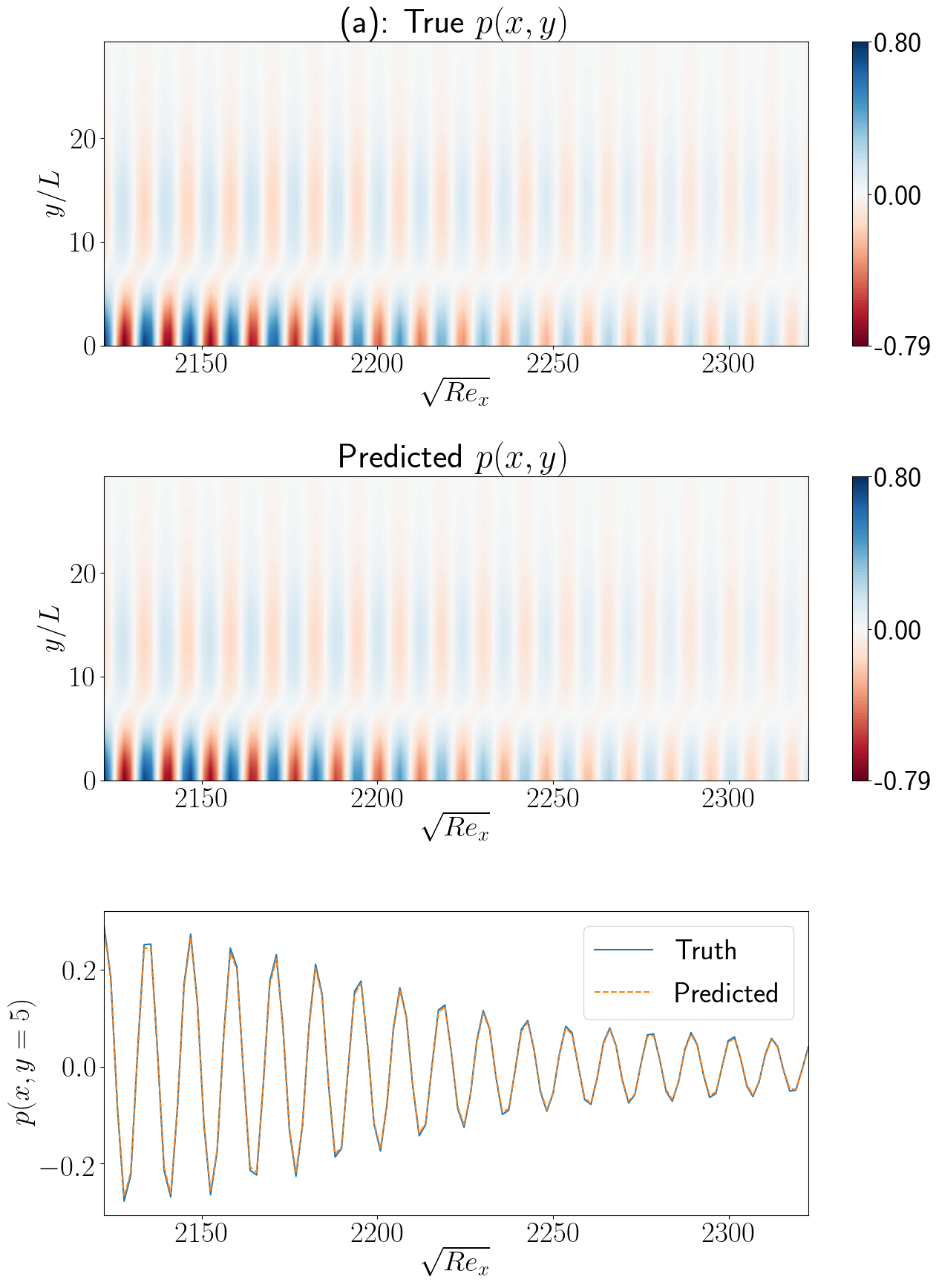}
    \includegraphics[width=0.45\textwidth]{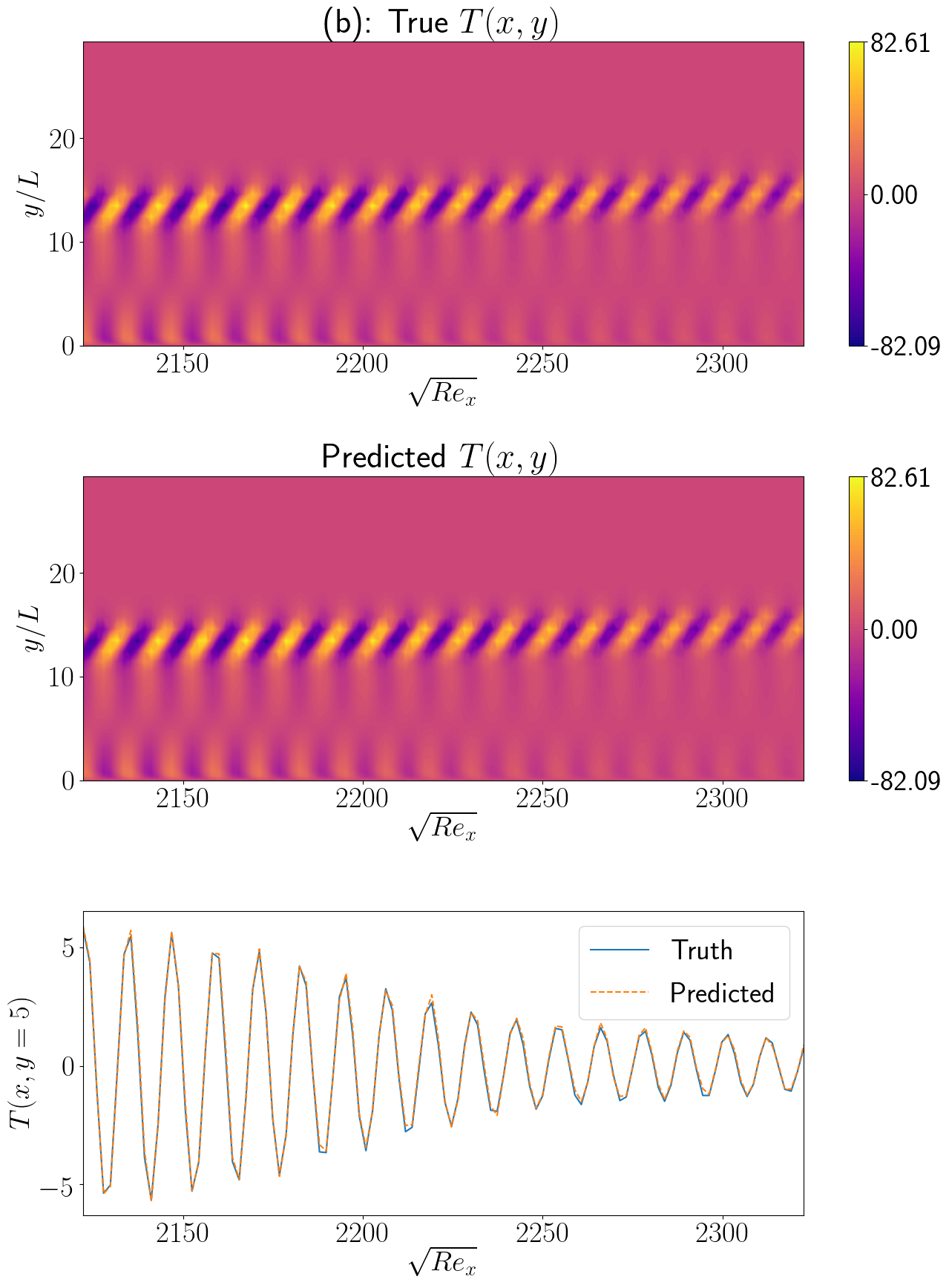}
    \caption{Examples from cases (a) F$_p$ and (b) F$_T$. For one particular input mode frequency, belonging to the validation dataset, we show the true solutions as generated by the PSE on the top row, the prediction obtained from the DeepONet on the middle row, and a comparison of the profiles for a fixed $y/L_0=5$ on the bottom row.}
    \label{fig:examples_FpandT}
\end{figure*}

\begin{figure}
    \includegraphics[width=0.40\textwidth]{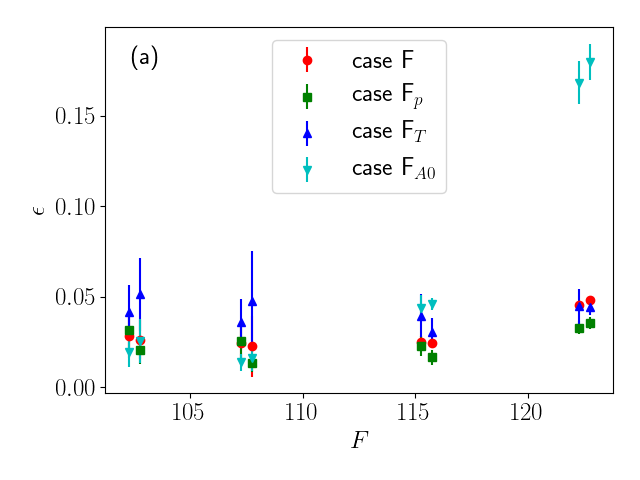}
    \includegraphics[width=0.40\textwidth]{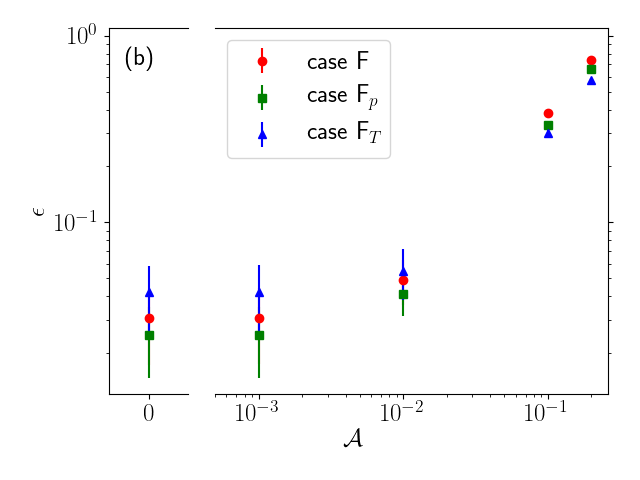}
    \caption{(a): Relative root mean square errors $\epsilon$ evaluated over the whole output domain for different solutions at the various frequencies present in the validation datasets for cases F, F$_p$, F$_T$ and F$_{A0}$. (b): Relative root mean square errors $\epsilon$ evaluated over the whole output domain for different solutions using noisy data as a function of the noise amplitude $\mathcal{A}$ for cases F, F$_p$ and F$_T$.}
    \label{fig:errs_forward}
\end{figure}

For a quantitative assessment of the performance of the DeepONet, we define $\epsilon$ as the relative root mean square error evaluated over the full output domain for a given input mode,
\begin{equation}
    \epsilon(f) = \sqrt{\frac{\langle \left[G(f)(\zeta) - G^\dagger(f)(\zeta) \right]^2 \rangle_\zeta}{\langle [G^\dagger(f)(\zeta)]^2 \rangle_\zeta}},
\end{equation}
where the average operation $\langle\cdot\rangle_\zeta$ is performed over the output domain $Y$. We calculated $\epsilon$ for a number of randomly selected modes out of the validation set.  The mean and standard deviation of $\epsilon$ are shown in Figure~\ref{fig:errs_forward}(a), grouped by the frequency $F$ of the input mode. The errors for all modes in the validation set are below 5\%.

We evaluated the robustness of the trained DeepONet against noisy input data to the branch. We introduced an additive white noise term scaled by the amplitude of the respective input function $f$, i.e.,
\begin{equation}
    f \mapsto f + \mathcal{A} \max{f} \eta,    
\end{equation}
where $\eta$ is a random Gaussian correlated field and $\mathcal{A}$ is the effective noise amplitude. Then, $\epsilon$ is calculated by drawing modes from the validation dataset, similar to the process performed in Figure~\ref{fig:errs_forward}(a) but without separating the results into the different input frequencies. The prediction errors are shown in Figure~\ref{fig:errs_forward}(b).  The results demonstrate that the accuracy of DeepONet predictions are unchanged when the noise is up to 1\%, but deteriorates at higher noise amplitudes.  

The results from cases F$_p$ and F$_T$ are similar to those from case F. Sample predictions for the two configurations are shown in Figure~\ref{fig:examples_FpandT}, while the relative errors for the different modes in the validation dataset are shown in Figure~\ref{fig:errs_forward}(a). As with the previous case, the trained DeepONets are able to accurately map inflow perturbations to the different fields downstream and can differentiate between the different modal profiles, frequencies, and spatially dependent growth rates.


\subsection{Computational cost and proposed metrics for DeepONet characterization}
 
The preceding cases each require approximately 3{,}000 minutes to train using an Nvidia Tesla K80 GPU card. Evaluating the DeepONet with one batch of $5{,}000$ points (roughly the number of points used to represent the field in $Y_d$) entails on the order of $10^8$ floating point operations and takes $2.5\times 10^{-2}$ seconds using the same card. In order to view these computational requirements in context, we can compare them to costs associated to the PSE solution. We stress, however, that DeepONets cannot be regarded as replacement for classical numerical simulations which are needed anyway to generate data.
The PSE solution was performed on an Intel Core i5 CPU and required approximately 15 minutes; 59 different modes were evaluated to generate the training and validation data, thus totalling 885 minutes of single CPU time.

We define the following three metrics: the training ratio, $R_t$, the evaluation ratio, $R_e$, and break even number, $N^*_e$, as
\begin{equation}
    R_t = \frac{C_t}{N_s C_s},
    \qquad
    R_e =  \frac{C_e}{C_s},
    \qquad
    N^*_e = N_s + \frac{C_t}{C_s},
\end{equation}
where $C_t$ is the cost in time of training the DeepONet, $N_s$ is the number of simulations needed to generate the dataset, $C_s$ is the cost in time of running each simulation, and $C_e$ is cost in time of evaluating  a DeepONet. The training ratio compares the cost of training against the cost of generating data. The evaluation ratio compares the cost of evaluating the DeepONet against the cost of performing a PSE solution. The break even number indicates the number of evaluations at which the DeepONet becomes beneficial compared to the simulation tool, and stems from the analysis of the total cost ratio
\begin{equation}
   R_c =  \frac{N_s C_s + C_t + N_e C_e}{N_e C_s},
\end{equation}
which compares the total cost of generating data, training the DeepONet and evaluating $N_e$ different solutions to the cost of generating all $N_e$ different solutions with simulations. Equating the ratio $R_c$ to unity and using that $C_e \ll C_s$ yields the expression for $N^*_e$ reported above.

For our results, we obtain $R_t = 3.39$, $R_e = 2.7\times 10{-5}$, and $N^*_e=250$.
These numbers show that the training time of a DeepONet is manageable and comparable to the data generation in the present case and, as expected, that evaluating a trained DeepONet is a very fast operation. It is also important to state that these values may vary strongly depending on the particular application.

\subsection{Impact of loss function weighting and feature expansion}

\begin{figure}[htbp]
    \includegraphics[width=0.40\textwidth]{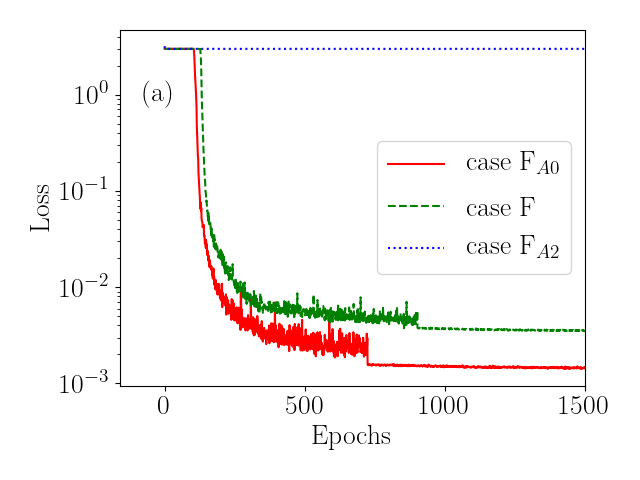}
    \includegraphics[width=0.40\textwidth]{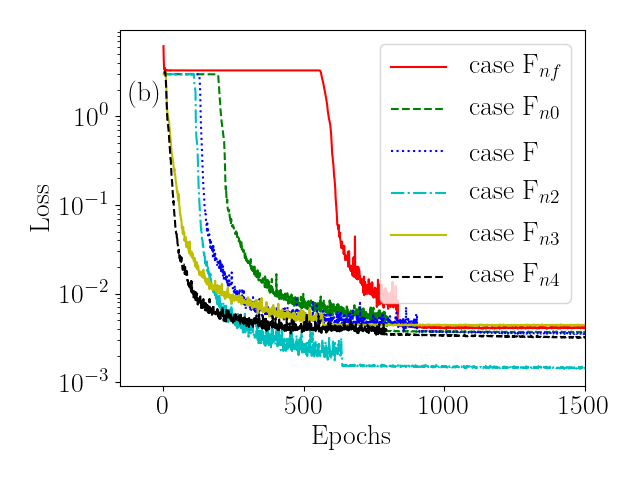}
    \caption{(a): Evolution of the loss function evaluated on the validation datasets for cases F, F$_{A0}$ and F$_{A2}$.
    (b): Evolution of the loss function evaluated on the validation datasets for cases F$_{nf}$, F$_{n0}$, F, F$_{n2}$, F$_{n3}$ and F$_{n4}$.}
    \label{fig:loss_impact}
\end{figure}

Some of the algorithmic considerations when training DeepONets are important to note. The first one is the use of weights in the loss function. While all frequencies studied are unstable at the inflow to the flow domain, the higher frequencies decay considerably as they evolve  downstream. As a result, the target fields differ in amplitudes by more than two orders of magnitude. Having such disparity in the dataset poses a problem for the gradient descent-based training protocol. The gradient of Eq.~\eqref{eq:loss_deeponet} with respect to the network parameters $\bm{\theta}$ takes the form
\begin{equation}
    \frac{\partial L}{\partial \bm{\theta}} = \frac{2}{N} \sum_{i=1}^N w_i ( G(f_i)(\zeta_i) - G^\dagger(f_i)(\zeta_i) ) \frac{\partial G(f_i)(\zeta_i)}{\partial \bm{\theta}}.
    \label{eq:loss_deeponet}
\end{equation}
During training $G_i - G^\dagger_i \propto A_i$ (with $A_i$ defined in Eq.~\eqref{eq:weights}), and thus solutions that have low output amplitude produce smaller changes to the loss function when applying a gradient update than those with high output amplitude, unless $w_i$ is used to balance the problem. In Figure~\ref{fig:loss_impact}(a) we show the evolution of the loss function evaluated on the validation dataset for cases F$_{A0}$, F and F$_{A2}$, which used $w_i$ equal to $1$, $A^{-1}_i$ and $A^{-2}_i$, respectively. All three cases go through a plateau where networks output the mean, and the overcompensated case $F_{A2}$ is not able to exit this state during the number of epochs shown. The non-compensated case $F_{A0}$ and compensated case $F$ are able to learn the solutions of the PSE. While the former reaches a lower loss, its prediction errors are not necessarily smaller. The values of $\epsilon$ grouped by frequency from both cases $F_{A0}$ and $F$ are shown in Fig.~\ref{fig:errs_forward}, indicating that not compensating for the differences in amplitudes limits the ability of the former network to correctly learn the low-amplitudes modes.

The existence of a plateau during training represents an extra cost for DeepONets. Expanding input features using Eq.~\eqref{eq:feat1} proved to be a key element in reducing the time that the networks spend in the plateau. The values of the loss function evaluated over the validation dataset for cases F$_{nf}$, F$_{n0}$, F, F$_{n2}$, F$_{n3}$ and F$_{n4}$ are shown in Figure~\ref{fig:loss_impact}(b). The feature expansion reduces the duration of the plateau during training, to the point of omitting it completely when using a large number of modes ($n=3$ and 4). All cases converge to the same values, expect for case F$_{n2}$ which converges to slightly lower value. To the best of our knowledge, there is nothing special in relation to our data about using $n=2$. We note that we did not perform an exhaustive optimization of the hyperparameters of the networks used, so the possibility exists of improving our results through hyperparamenter tuning.



\subsection{Inverse problem}

\begin{figure}[htbp]
    \includegraphics[width=0.40\textwidth]{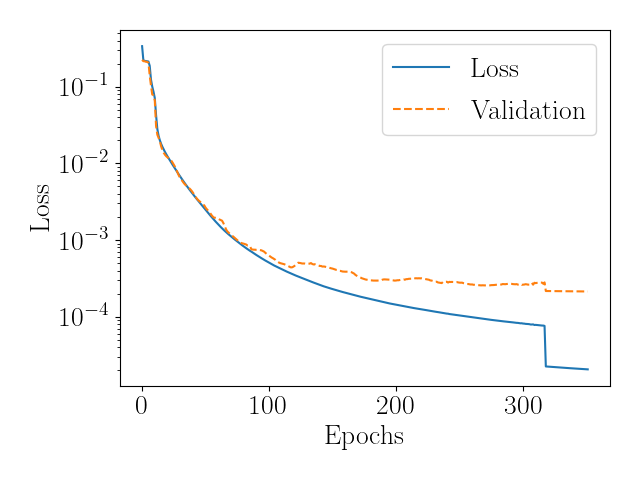}
    \caption{Evolution of the loss function evaluated on the training and validation datasets for case I.}
    \label{fig:loss_I}
\end{figure}

\begin{figure}[htbp]
    \includegraphics[width=0.45\textwidth]{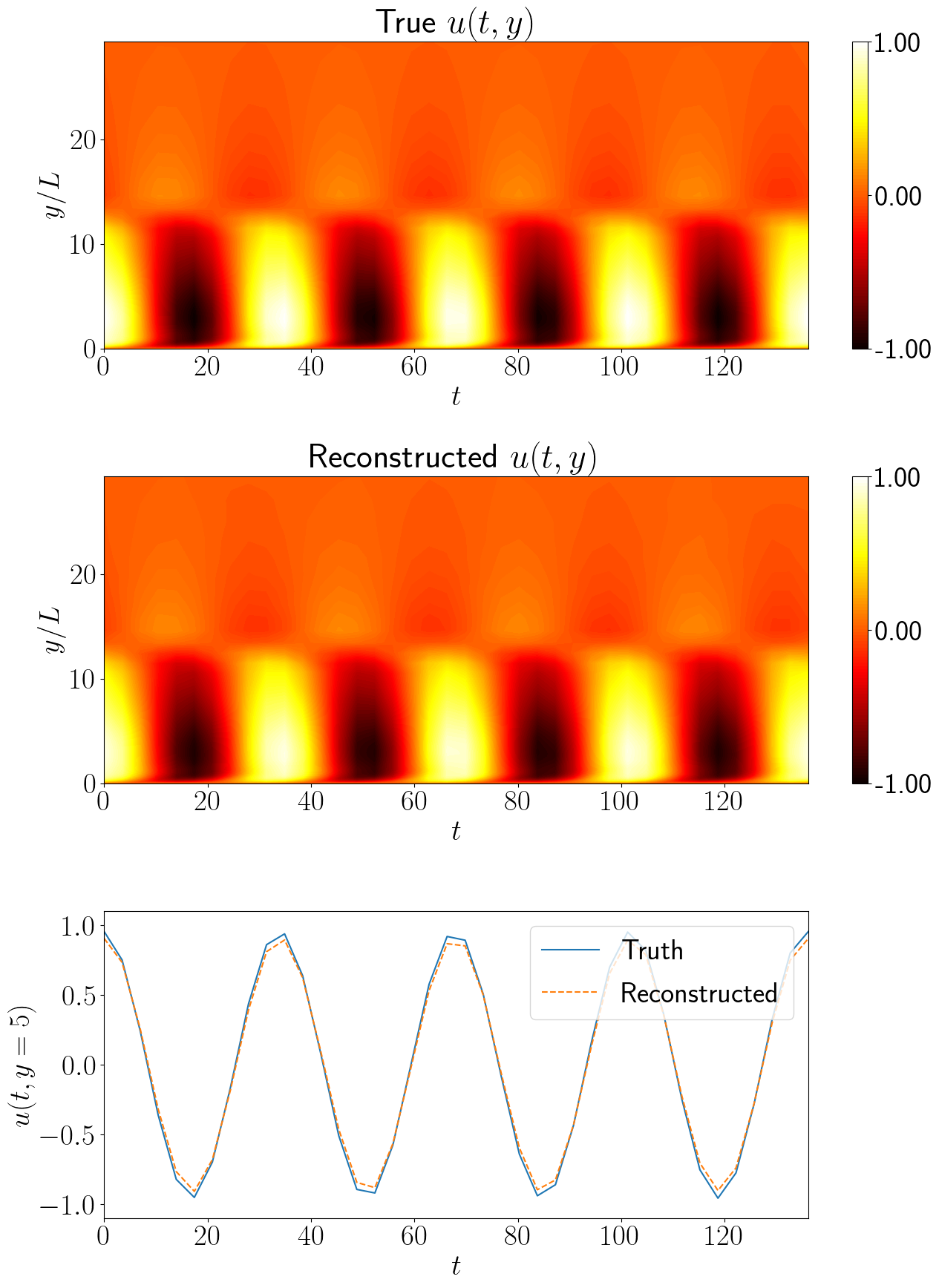}
    \caption{Example from case I. For one particular downstream pressure measurement, belonging to the validation dataset, we show the true upstream perturbation on the top row, the reconstruction obtained from the DeepONet on the middle row, and a comparison of the profiles for a fixed $y/L_0=5$ on the bottom row.}
    \label{fig:examples_I}
\end{figure}

\begin{figure}[htbp]
    \includegraphics[width=0.40\textwidth]{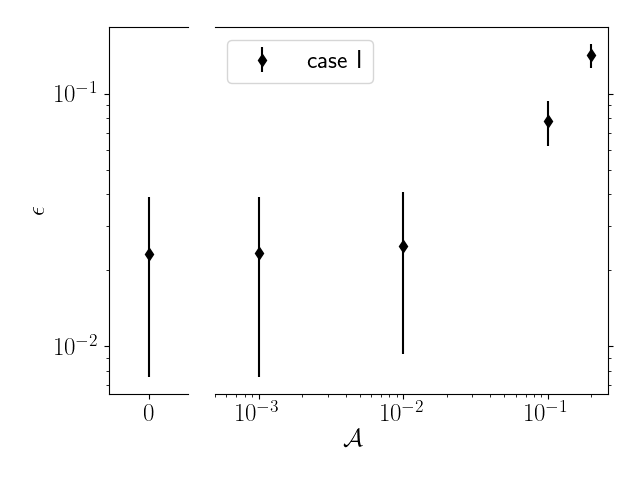}
    \caption{Relative root mean square errors $\epsilon$ evaluated over the whole output domain for different solutions using noisy data as a function of the noise amplitude $\mathcal{A}$ for case I.}
    \label{fig:errs_noise_inverse}
\end{figure}

The evolution of the loss functions for both the training and validation datasets for case I is shown in Figure~\ref{fig:loss_I}. The network is able to learn how to map from downstream measurements of the wall pressure to 
inflow perturbation modes. Contrary to the forward cases, no plateau arises during training nor did the loss function require weighting, as the outputs of the network are all of the same order of magnitude. An example of a reconstructed inflow mode evaluated over the whole output domain is shown in Figure~\ref{fig:examples_I}. The overall accuracy and robustness with respect to noisy inputs are analyzed in Figure~\ref{fig:errs_noise_inverse}, where the relative error $\epsilon$ is plotted as function of the input noise amplitude $\mathcal{A}$. The inverse case yields results similar or better than the forward cases.

\subsection{Two-mode cases}

\begin{figure}[htbp]
    \includegraphics[width=0.40\textwidth]{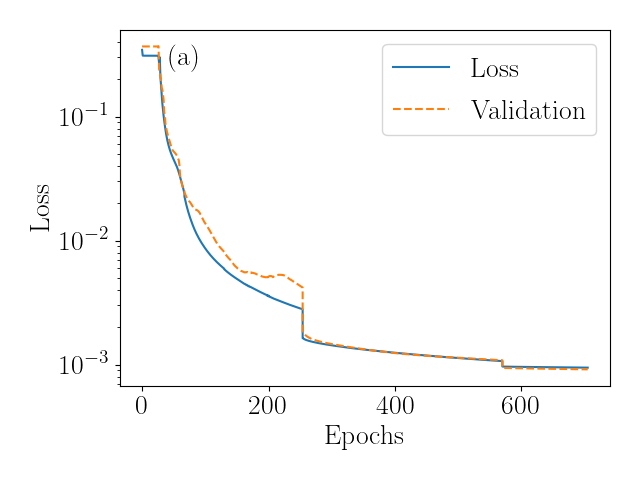}
    \includegraphics[width=0.40\textwidth]{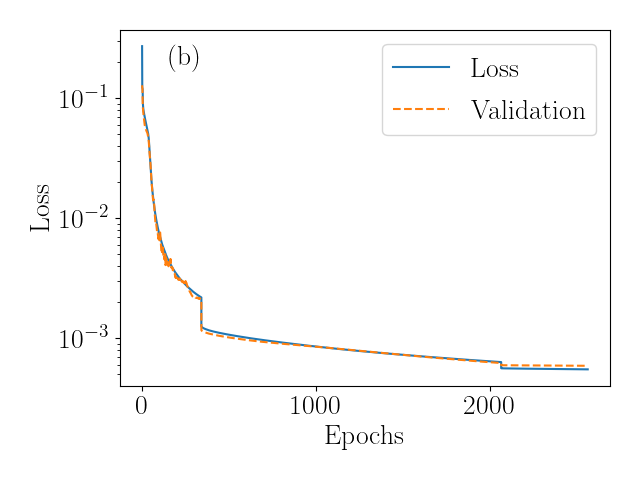}
    \caption{Evolution of the loss function evaluated on the training and validation datasets for cases (a) F$_2$ and (b) I$_2$.}
    \label{fig:loss_twomodes}
\end{figure}

\begin{figure}[htbp]
    \includegraphics[width=0.45\textwidth]{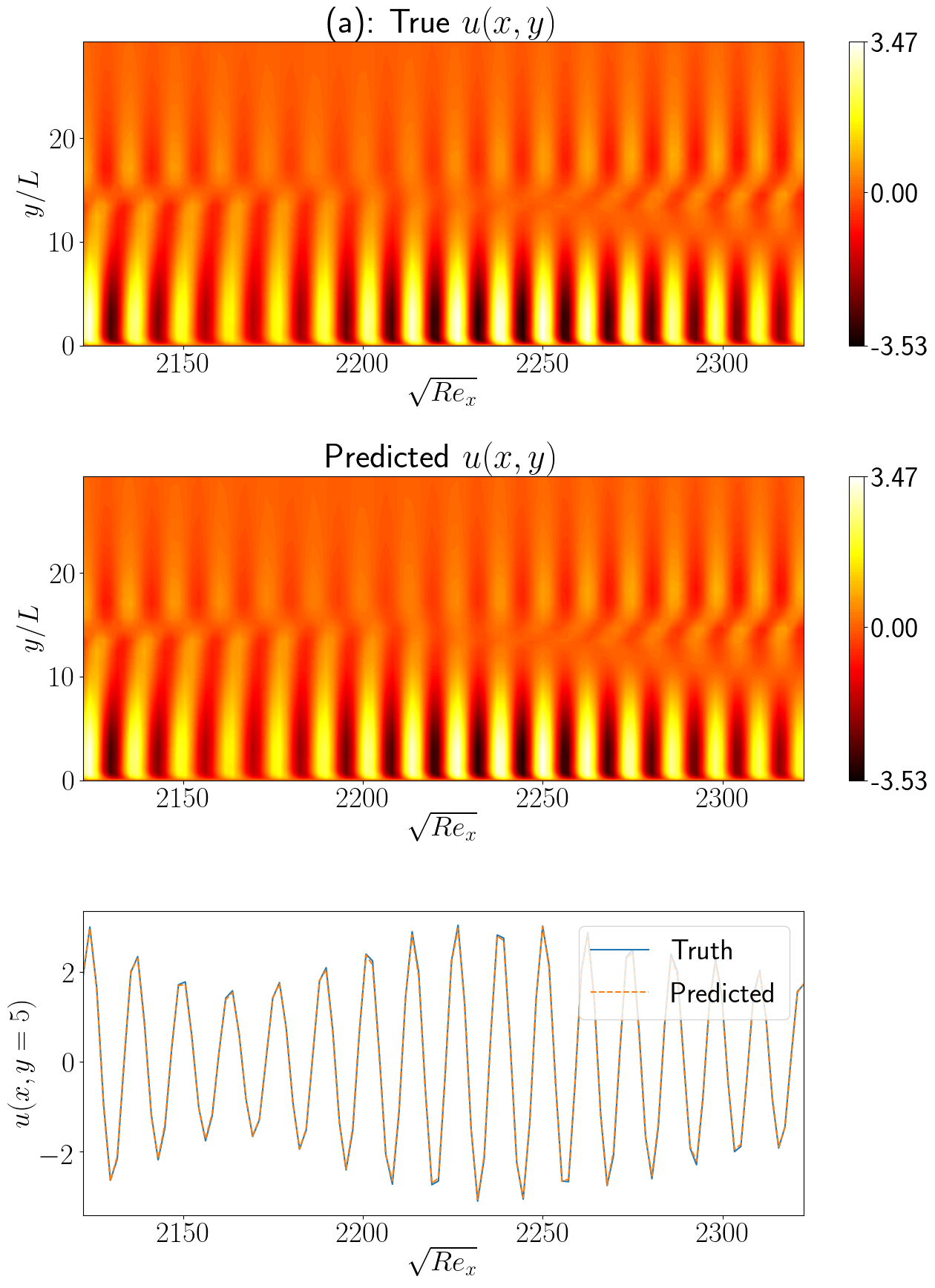}
    \includegraphics[width=0.45\textwidth]{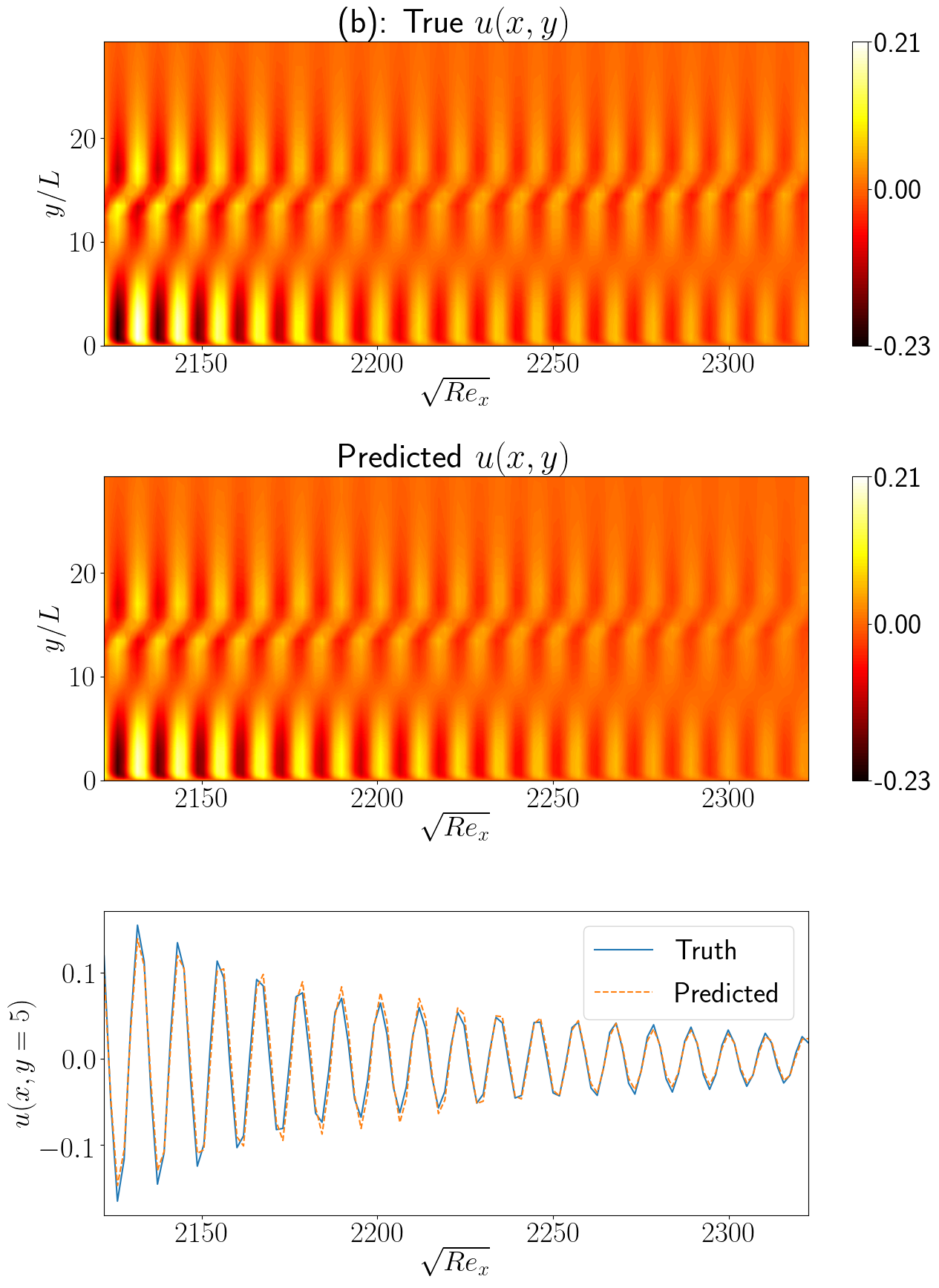}
    \caption{Examples from case F$_2$. For two particular input mode combinations, one in panel (a) and another in panel (b) with both belonging to the validation dataset, we show the true solutions as generation from the PSE on the top row, the prediction obtained from the DeepONet on the middle row, and a comparison of the profiles for a fixed $y/L_0=5$ on the bottom row.}
    \label{fig:examples_F2}
\end{figure}

\begin{figure}[htbp]
    \includegraphics[width=0.45\textwidth]{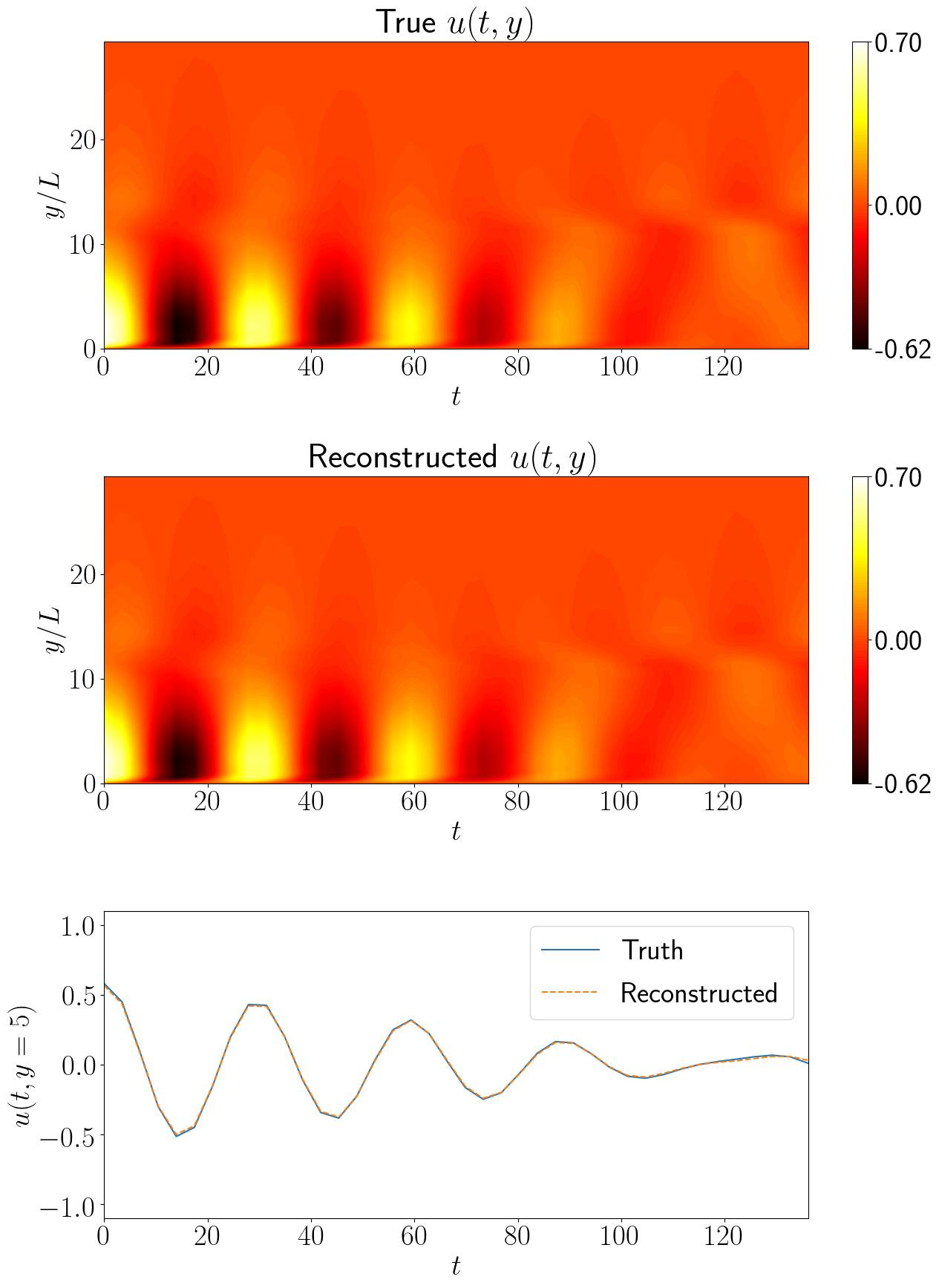}
    \caption{Example from case I$_2$. For one particular downstream pressure measurement, belonging to the validation dataset, we show the true upstream perturbation on the top row, the reconstruction obtained from the DeepONet on the middle row, and a comparison of the profiles for a fixed $y/L_0=5$ on the bottom row.}
    \label{fig:examples_I2}
\end{figure}

The next cases we analyze are F$_2$ and I$_2$, where the DeepONets were trained to learn linear combinations of solutions. The loss functions are shown in Figure~\ref{fig:loss_twomodes} and  examples of prediction and reconstruction are shown in Figures~\ref{fig:examples_F2} and \ref{fig:examples_I2}, for case F$_2$ and I$_2$, respectively.  Once again, the DeepONets are able to learn their target solutions. However, training is more challenging because the effective solution space is much larger than for a single instability mode; here the perturbation is comprised of two instability waves with different frequencies, phases and amplitudes. This is evidenced in Figure~\ref{fig:examples_F2}(b) which shows very good qualitative agreement between data and prediction, but has a relative error $\epsilon=0.185$ that is appreciably higher than the  values encountered in the earlier cases.

\begin{figure}
    \includegraphics[width=0.40\textwidth]{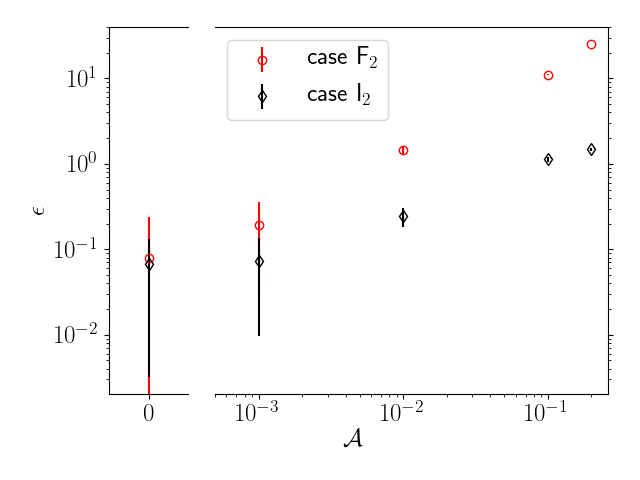}
    \caption{Relative root mean square errors $\epsilon$ evaluated over the whole output domain for different solutions using noisy data as a function of the noise amplitude $\mathcal{A}$ for cases F$_2$ and I$_2$.}
    \label{fig:errs_noise_twomodes}
\end{figure}

We analyze the robustness to input uncertainties in Fig.~\ref{fig:errs_noise_twomodes}. As expected from the comment above, the average $\epsilon$ when $\mathcal{A}=0$ is higher than in the respective cases where only one mode was used to generate the solutions, and the increased difficulties in training also lead to the networks being more sensitive to noise. Nonetheless, the predictions and reconstructions generated  by the DeepONets are still satisfactory. 

Due to its complexity, case F$_2$ was chosen for comparison of DeepONets against CNNs. The results are shown in Appendix~\ref{app:cnns}. DeepONets achieve slightly better results than CNNs but at an increased training cost. In general, while CNNs are well suited for problems that can be cast onto a rectangular grid, DeepONets are more flexible and can deal with data of different shapes in both input and output. DeepONets can also make predictions at arbitrary locations, contrary to the fixed output of a CNN, and are thus then capable of utilizing physics-informed constraints \cite{wang_learning_2021}, as it is possible to apply automatic differentiation to the trunk input variables, or use Fourier feature expansion, as shown above. 

\subsection{Data Assimilation using trained DeepONets}

\begin{figure}[htbp]
    \includegraphics[width=0.40\textwidth]{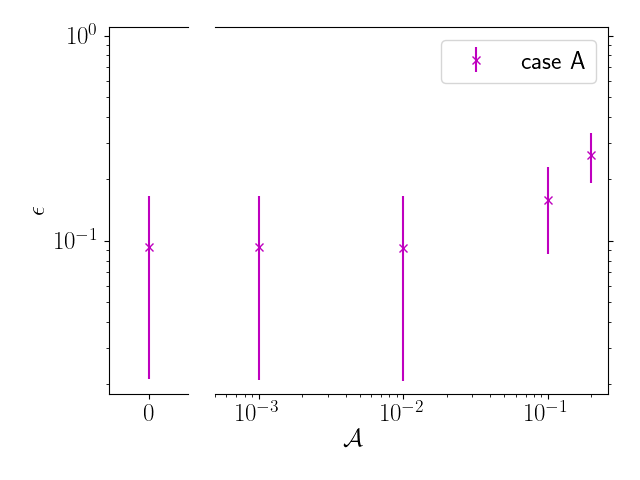}
    \caption{Relative root mean square errors $\epsilon$ evaluated over the whole output domain for different solutions using noisy data as a function of the noise amplitude $\mathcal{A}$ for case A.}
    \label{fig:errs_noise_da}
\end{figure}

Finally, we analyze a prototype for data assimilation (DA) using DeepONets. While DeepONets could be trained to directly map measurements to flow-field prediction,  we look at the case where we concatenate two already trained networks. In particular, the output of case I is fed into case F and we term this configuration case A. Figure~\ref{fig:errs_noise_da} shows the value of $\epsilon$ at different noise levels for this case. The proposed DA protocol is able to reconstruct the inflow condition and predict the corresponding downstream field. The higher errors obtained when $\mathcal{A}=0$ are due to the errors present in the output of case I.

\begin{figure*}
    \includegraphics[width=0.45\textwidth]{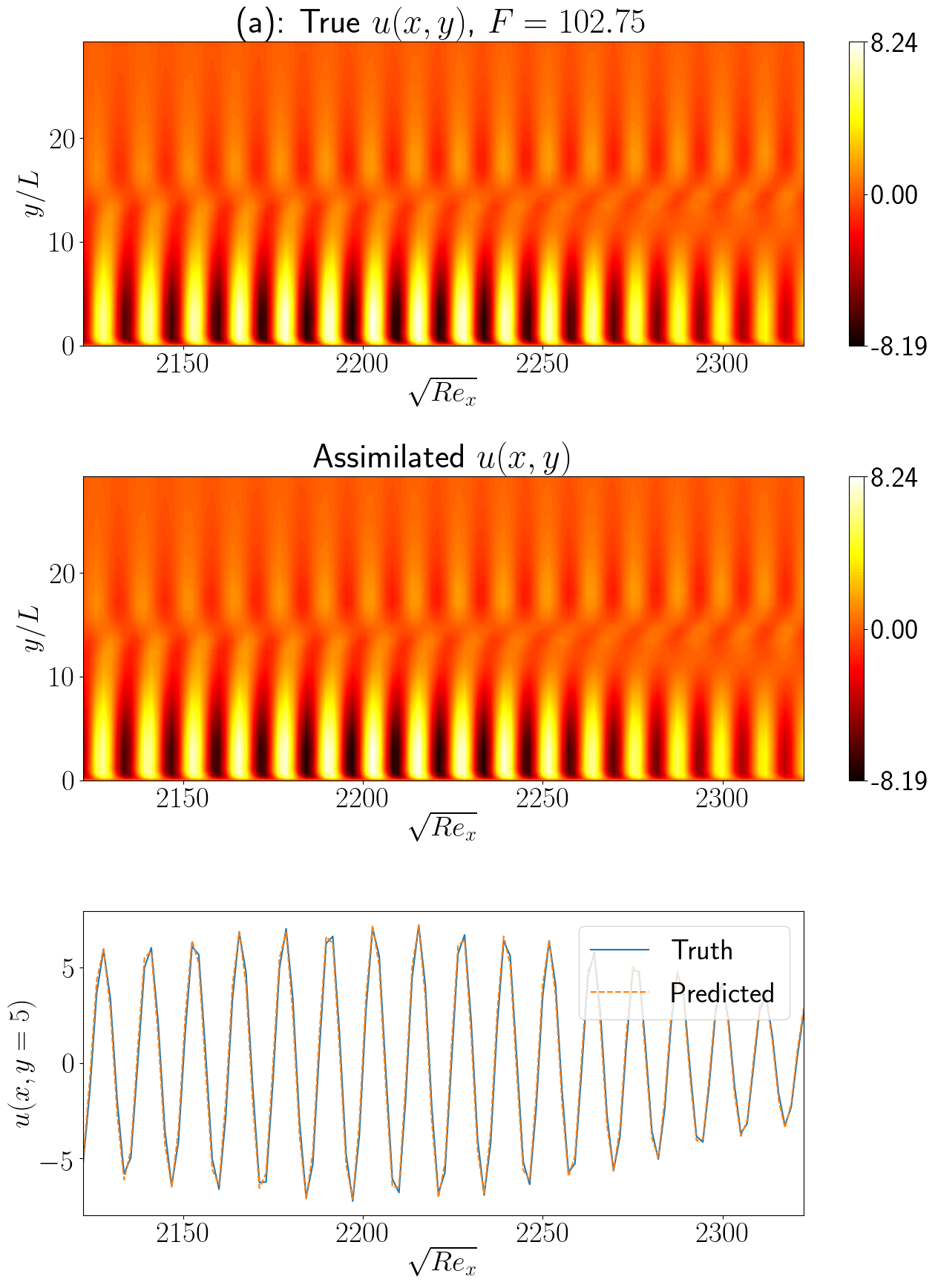}
    \includegraphics[width=0.45\textwidth]{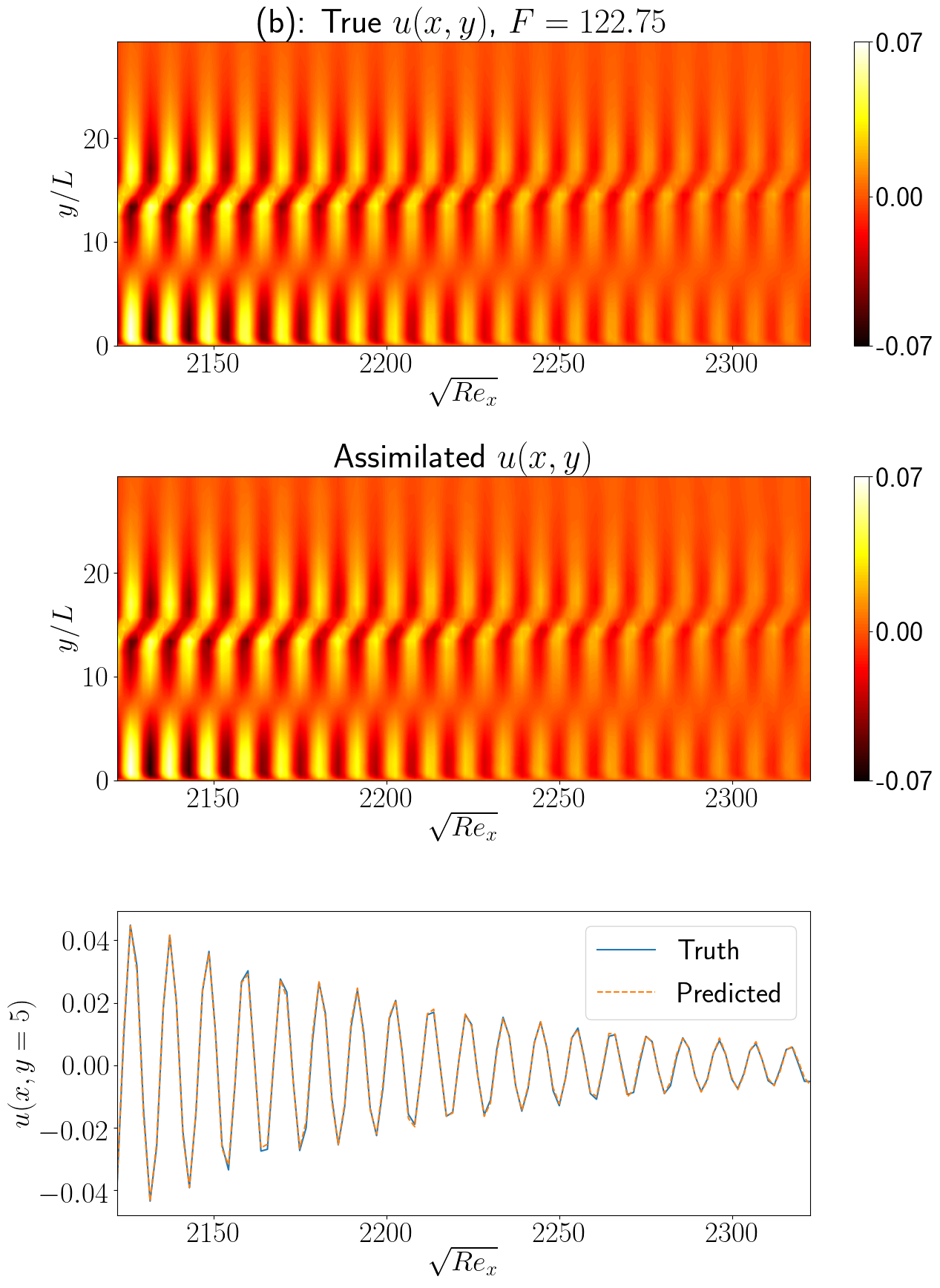}
    \caption{Examples from case A. For two particular downstream pressure measurements, (a) $F=102.75$ and (b) $F=122.75$, we show the true solutions as generated from the PSE on the top row, the assimilated fields obtained from the DeepONet on the middle row, and a comparison of the profiles for a fixed $y/L_0=5$ on the bottom row.}
    \label{fig:examples_A}
\end{figure*}

In Figure~\ref{fig:examples_A} we show two examples of assimilated fields, using the same solutions as in Figure~\ref{fig:examples_F}. Qualitatively, cases F and A produce very similar results, but the errors are slightly higher for case A. The error $\epsilon$ for the example shown in Figure~\ref{fig:examples_A}(a) is equal to $0.112$, compared to $\epsilon=0.030$ as obtained for Figure~\ref{fig:examples_F}(a). For the example shown in panel (b) of both figures, $\epsilon=0.047$ when evaluating directly, only increasing slightly to $\epsilon=0.054$ when performing the assimilation.
Similar to what was shown in Figure~\ref{fig:examples_F2}(b), the main sources of errors are slight shifts in phase and amplitude.
Thanks to the robustness of case I to input uncertainty, case A scales slightly better than case F. Overall, DeepONets can act as efficient and flexible DA frameworks.

\section{Conclusions}
\label{sec:conclusions}

Deep learning techniques have made great strides in numerous problems in computer science, and their capabilities for important applications in physics and engineering are expanding.  Generating fast and accurate solutions of systems of equations is one such major problem where machine learning is posed to accelerate scientific discovery.  We showed that DeepONets, which are neural network architectures designed to approximate not just functions but operators, can map an upstream instability wave in a high Mach number boundary layer to a downstream field. We introduced three different cost metrics as a way to assess the feasibility of DeepONets for the application at hand. DeepONets can predict different components of the state vector, and also map downstream measurements to upstream disturbances, which is beneficial for inverse modeling and data assimilation. The introduction of Fourier harmonic input feature expansion and loss function weighting were key elements to speed up of training and achieve accurate predictions. DeepONets were also shown to be robust against noisy inputs and be able to perform data assimilation, even though they were not explicitly trained for either task. Improvements to the training procedure that take into account noisy data can further improve the performance of DeepONets.

\appendix

\section{Operator matrices of the Parabolized Stability Equations}
\label{appendix:matrix}

We present the different  non-zero elements of the operator matrices $\mathcal{V}$, featured in Eq.~\eqref{eq:lin_ptb_eq} and operator $\check{\mathcal{A}}$, used in Eq.~\eqref{eq:pse}. Note that terms with order $O({1}/{Re_{0}^{2}})$ are neglected \cite{Bertolotti1991}.  The indices $i$, $j$ represent the row and the column entries within the matrix operator:

\begin{equation}
\mathcal{V}_{t}(1,1)=1,~\mathcal{V}_{t}(2,2)=\mathcal{V}_{t}(3,3)=\mathcal{V}_{t}(4,4)=\rho_{B},
\end{equation}
\begin{equation}
\mathcal{V}_{0}(1,1)=\frac{\partial U_{B}}{\partial x}+\frac{\partial V_{B}}{\partial y},~\mathcal{V}_{0}(1,2)=\frac{\partial \rho_{B}}{\partial x},
\end{equation}
\begin{equation}
\mathcal{V}_{0}(2,1)=U_{B}\frac{\partial U_{B}}{\partial x}+V_{B}\frac{\partial U_{B}}{\partial y}+\frac{1}{\gamma_{0} M_{0}^{2}}\frac{\partial T_{B}}{\partial x},~\mathcal{V}_{0}(2,2)=\rho_{B}\frac{\partial U_{B}}{\partial x},
\end{equation}

\begin{equation}
\begin{aligned}
\mathcal{V}_{0}(2,4)
&=\frac{1}{\gamma_{0} M_{0}^{2}}\frac{\partial \rho_{B}}{\partial x} -\frac{1}{Re_{0}}\left[ l\left\{ \left(\frac{\partial^{2}U_{B}}{\partial x^{2}}+\frac{\partial^{2}V_{B}}{\partial x\partial y}\right)\frac{\partial\mu_{B}}{\partial T_{B}} \right. \right.\\
&\left. \left.+\left(\frac{\partial U_{B}}{\partial x}+\frac{\partial V_{B}}{\partial y}\right)\frac{\partial^{2}\mu_{B}}{\partial T_{B}^{2}}\frac{\partial T_{B}}{\partial x}\right\}\right.\\
&+2\left(\frac{\partial^{2}U_{B}}{\partial x^{2}}\frac{\partial\mu_{B}}{\partial T_{B}}+\frac{\partial U_{B}}{\partial x}\frac{\partial^{2}\mu_{B}}{\partial T_{B}^{2}}\frac{\partial T_{B}}{\partial x}\right)\\
&+\left(\frac{\partial^{2}U_{B}}{\partial y^{2}}+\frac{\partial^{2}V_{B}}{\partial x\partial y}\right)\frac{\partial\mu_{B}}{\partial T_{B}}\\
&\left.+\left(\frac{\partial U_{B}}{\partial y}+\frac{\partial V_{B}}{\partial x}\right)\frac{\partial^{2}\mu_{B}}{\partial T_{B}^{2}}\frac{\partial T_{B}}{\partial y}\right],
\end{aligned}
\end{equation}

\begin{equation}
\mathcal{V}_{0}(4,1)=U_{B}\frac{\partial T_{B}}{\partial x}+V_{B}\frac{\partial T_{B}}{\partial y}+(\gamma_{0}-1)T_{B}\left(\frac{\partial U_{B}}{\partial x}+\frac{\partial V_{B}}{\partial y}\right),
\end{equation}

\begin{equation}
\mathcal{V}_{0}(4,2)=\rho_{B}\frac{\partial T_{B}}{\partial x},
\end{equation}

\begin{equation}
\begin{aligned}
\mathcal{V}_{0}(4,4)&=(\gamma_{0}-1)\rho_{B}\left(\frac{\partial U_{B}}{\partial x}+\frac{\partial V_{B}}{\partial y}\right)\\
&-\frac{\gamma_{0} }{Re_{0}Pr_{0}}\left[\left(\frac{\partial^{2}T_{B}}{\partial x^{2}}+\frac{\partial^{2}T_{B}}{\partial y^{2}}\right)\frac{\partial k_{B}}{\partial T_{B}}\right.\\
&+\left\{\left(\frac{\partial T_{B}}{\partial x}\right)^{2}\right. \left.\left.+\left(\frac{\partial T_{B}}{\partial y}\right)^{2}\right\}\frac{\partial^{2}k_{B}}{\partial T_{B}^{2}}\right]\\
&-\frac{\gamma_{0}(\gamma_{0}-1)M_{0}^{2}}{Re_{0}}\frac{\partial\mu_{B}}{\partial T_{B}}\left[2\left\{\left(\frac{\partial U_{B}}{\partial x}\right)^{2}+\left(\frac{\partial V_{B}}{\partial y}\right)^{2}\right\}\right. \\
& \left.+\left(\frac{\partial V_{B}}{\partial x}+\frac{\partial U_{B}}{\partial y}\right)^{2}+l\left(\frac{\partial U_{B}}{\partial x}+\frac{\partial V_{B}}{\partial y}\right)^{2}\right],
\end{aligned}
\end{equation}
\begin{equation}
\mathcal{V}_{x}(1,1)=U_{B},~\mathcal{V}_{x}(1,2)=\rho_{B},
\end{equation}
\begin{equation}
\mathcal{V}_{x}(2,1)=\frac{T_{B}}{\gamma_{0} M_{0}^{2}},~\mathcal{V}_{x}(2,2)=\rho_{B}U_{B}-\frac{l+2}{Re_{0}}\frac{\partial\mu_{B}}{\partial T_{B}}\frac{\partial T_{B}}{\partial x},
\end{equation}

\begin{equation}
\mathcal{V}_{x}(2,4)=\frac{\rho_{B}}{\gamma_{0} M_{0}^{2}}-\frac{1}{Re_{0}}\frac{\partial\mu_{B}}{\partial T_{B}}\left[l\left(\frac{\partial U_{B}}{\partial x}+\frac{\partial V_{B}}{\partial y}\right)+2\frac{\partial U_{B}}{\partial x}\right],
\end{equation}
\begin{equation}
\mathcal{V}_{x}(3,3)=\rho_{B}U_{B}-\frac{1}{Re_{0}}\frac{\partial\mu_{B}}{\partial T_{B}}\frac{\partial T_{B}}{\partial x},
\end{equation}
\begin{equation}
\mathcal{V}_{x}(4,2)=(\gamma_{0}-1)-\frac{2\gamma_{0}(\gamma_{0}-1)M_{0}^{2}\mu_{B}}{Re_{0}}\left[(l+2)\frac{\partial U_{B}}{\partial x}+l\frac{\partial V_{B}}{\partial y}\right],
\end{equation}
\begin{equation}
\mathcal{V}_{x}(4,4)=\rho_{B}U_{B}-\frac{2\gamma_{0}}{Re_{0}Pr_{0}}\frac{\partial k_{B}}{\partial T_{B}}\frac{\partial T_{B}}{\partial x},
\end{equation}
\begin{equation}
\mathcal{V}_{y}(1,1)=V_{B},
\end{equation}
\begin{equation}
\mathcal{V}_{y}(2,2)=\rho_{B}V_{B}-\frac{1}{Re_{0}}\frac{\partial\mu_{B}}{\partial T_{B}}\frac{\partial T_{B}}{\partial y},
\end{equation}
\begin{equation}
\mathcal{V}_{y}(2,4)=-\frac{1}{Re_{0}}\frac{\partial\mu_{B}}{\partial T_{B}}\left(\frac{\partial U_{B}}{\partial y}+\frac{\partial V_{B}}{\partial x}\right),
\end{equation}
\begin{equation}
\mathcal{V}_{y}(3,3)=\rho_{B}V_{B}-\frac{1}{Re_{0}}\frac{\partial\mu_{B}}{\partial T_{B}}\frac{\partial T_{B}}{\partial y},
\end{equation}
\begin{equation}
\mathcal{V}_{y}(4,2)=-\frac{2\gamma_{0}(\gamma_{0}-1)M_{0}^{2}\mu_{B}}{Re_{0}}\left(\frac{\partial V_{B}}{\partial x}+\frac{\partial U_{B}}{\partial y}\right),\nonumber\\
\end{equation}
\begin{equation}
\mathcal{V}_{y}(4,4)=\rho_{B}V_{B}-\frac{2\gamma_{0}}{Re_{0}Pr_{0}}\frac{\partial k_{B}}{\partial T_{B}}\frac{\partial T_{B}}{\partial y},
\end{equation}
\begin{equation}
\mathcal{V}_{xx}(2,2)=-(l+2)\frac{\mu_{B}}{Re_{0}},
\end{equation}
\begin{equation}
\mathcal{V}_{xx}(3,3)=-\frac{\mu_{B}}{Re_{0}},~\mathcal{V}_{xx}(4,4)=-\frac{\gamma_{0}k_{B}}{Re_{0}Pr_{0}},
\end{equation}
\begin{equation}
\mathcal{V}_{yy}(2,2)=-\frac{\mu_{B}}{Re_{0}},
\end{equation}
\begin{equation}
\mathcal{V}_{yy}(3,3)=-\frac{\mu_{B}}{Re_{0}},
\end{equation}
\begin{equation}
\mathcal{V}_{yy}(4,4)=-\frac{\gamma_{0}k_{B}}{Re_{0}Pr_{0}},
\end{equation}
%


%
\begin{align}
\check{\mathcal{A}}(1,1)&=U_{B},\\
\check{\mathcal{A}}(1,2)&=\rho_{B},\\
\check{\mathcal{A}}(2,2)&=\check{\mathcal{A}}(3,3)=\check{\mathcal{A}}(4,4)=\rho_{B}U_{B},\\
\check{\mathcal{A}}(2,1)&=\frac{T_{B}}{\gamma_{0}M_{0}^{2}},\\
\check{\mathcal{A}}(2,4)&=\frac{\rho_{B}}{\gamma_{0}M_{0}^{2}},\\
\check{\mathcal{A}}(4,2)&=(\gamma_{0}-1).
\end{align}

\section{Comparison between Convolutional Neural Networks and DeepONets}
\label{app:cnns}

\begin{figure}[htbp]
    \includegraphics[width=0.40\textwidth]{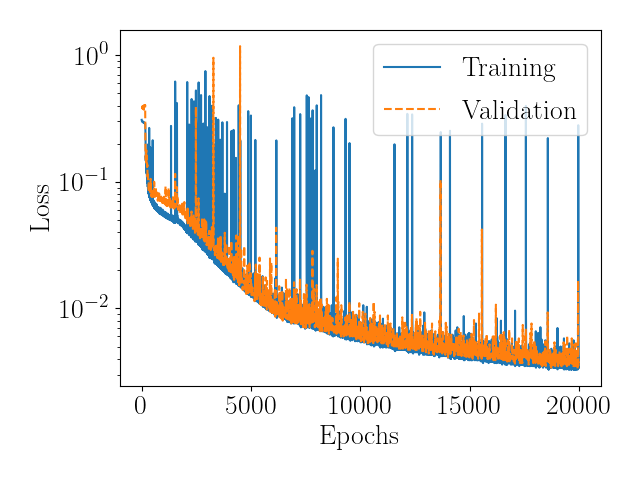}
    \caption{Evolution of the loss function evaluated on the training and validation datasets for case F$_2$ using a CNN-based encoder-decoder network.}
    \label{fig:loss_cnn}
\end{figure}

As with many problems typically addressed with deep learning tools, mapping inflow signals to downstream perturbations can be solved with other architectures besides DeepONets.  Convolutional neural networks (CNNs) are one possible and popular approach, which can be applied to our configuration. Here, we also train a CNN on the case F$_2$ dataset. Specifically, we first use a CNN-based encoder to map the input image to a low-dimensional latent space, and then use a CNN-based decoder to generate the output image from the latent space. We manually tuned the architecture hyperparameters (network size, regularization, batch size, learning rate, etc.), the smallest validation loss we obtained is $3.5\times 10^{-3}$, which is comparable to that of the DeepONet. In Figure~\ref{fig:loss_cnn} we show the evolution of the loss function. The values obtained are similar to those reported in Figure~\ref{fig:loss_twomodes}(a).

We expand on the differences between the two approaches and compare the results when applied to the same problem. CNNs have to work with gridded rectangular data both at the input and output, and require all the grid values during training. DeepONets can use arbitrary locations for their input which do not have to be a specific shape, and we can train DeepONets using partial observations. During the inference stage, DeepONets can evaluate their output at any location inside the domain, while CNNs can only predict the output on the grid. The flexibility of the DeepONet architecture also allows for the easy implementation of extra components, such as the feature expansion discussed in Sec.~\ref{sec:hyper} or even the adoption of convolutional layers in the branch network. Flexibility comes at a cost, however, as DeepONet training is less efficient than CNN. In the example shown above, performed using the same GPU in both cases, DeepONets were 7x times slower than the CNNs. The differences are listed in Table \ref{tab:deeponet_cnn}. Therefore, the only drawback of DeepONet is the high training cost, but we note the training is offline and this issue can also be relieved by data parallelism.

\begin{table}[htbp]
\begin{tabular}{l|cc}
\hline
 & DeepONet & CNN \\
\hline
Input/Output domain & Arbitrary & Rectangle \\
Mesh & Arbitrary & Grid \\
Training data & Partial observation & Complete observation \\
Prediction location & Arbitrary & Grid points \\
Architecture flexibility & \multicolumn{2}{c}{\begin{tabular}[c]{@{}c@{}}DeepONet is more flexible, e.g., adding features.
\end{tabular}} \\
Accuracy & \multicolumn{2}{c}{Comparable} \\
Training cost & \multicolumn{2}{c}{CNN is faster (up to 7X in our test).} \\
\hline
\end{tabular}
\caption{Comparison between DeepONets and CNNs.}
\label{tab:deeponet_cnn}
\end{table}

\bibliography{ref}

\end{document}